\documentclass[12pt, draftclsnofoot, onecolumn]{IEEEtran}

\usepackage{graphicx}
\usepackage{amsmath,dsfont,bbm,epsfig,amssymb,amsfonts, amstext, verbatim,amsopn,cite,subfigure,multirow,multicol,lipsum}
\usepackage{balance}
\usepackage{url}
\usepackage{amsfonts}
\usepackage{epsfig}
\usepackage{setspace}
\usepackage{stmaryrd}
\usepackage{psfrag}	
\usepackage{multirow}
\usepackage{float}
\usepackage[process=auto]{pstool}
\usepackage{etoolbox}
\usepackage{algorithmic}
\usepackage{algorithm}
\allowdisplaybreaks

\newtoggle{OneColumn}
\toggletrue{OneColumn}

\iftoggle{OneColumn}{%

}{%
}

\newcommand{\Equal}{\hspace{-0.4mm}=\hspace{-0.4mm}}

\newcommand{\Minus}{\hspace{-0.4mm}-\hspace{-0.4mm}}

\newcommand*{\Scale}[2][4]{\scalebox{#1}{$#2$}}%

\newtheorem{theo}{Theorem}

\newtheorem{remk}{Remark}

\EndPreamble
\begin{document}

\title{Optimal Relay Selection for the Parallel Hybrid RF/FSO Relay Channel:  \\  Non-Buffer-Aided and Buffer-Aided Designs
\vspace{-0.3cm}}
\author{Marzieh Najafi, Vahid Jamali, and Robert Schober \\
\IEEEauthorblockA{Friedrich-Alexander University (FAU), Erlangen, Germany}
\thanks{This paper was presented in part at IEEE ICC 2016 [14].}
\vspace{-0.5cm}}

\maketitle

\begin{abstract}
Hybrid radio frequency (RF)/free space optical (FSO) systems are among the candidate enabling technologies for the next generation of wireless networks since they benefit from both the high data rates of the FSO subsystem and the high reliability of the RF subsystem. In this paper, we focus on the problem of throughput maximization in the parallel hybrid RF/FSO relay channel. In the parallel hybrid RF/FSO relay channel, a source node sends its data to a destination node with the help of  multiple relay nodes. Thereby, for a given relay,  the source-relay and the relay-destination FSO links are orthogonal with respect to each other due to the narrow beam employed for FSO transmission, whereas, due to the broadcast nature of the RF channel,  half-duplex operation is required for the RF links if self-interference is to be avoided. Moreover, we consider the two cases where the relays are and are not equipped with buffers. For both cases, we derive the optimal relay selection policies for the RF and FSO links and the optimal time allocation policy for transmission and reception for the RF links.  The proposed optimal protocols provide important insights for optimal system design. Since the optimal buffer-aided (BA) policy introduces an unbounded end-to-end delay, we also propose a suboptimal BA policy which ensures certain target average delays. Moreover, we present distributed implementations for both proposed optimal protocols.  Simulation results demonstrate that a considerable gain can be achieved by the proposed adaptive protocols in comparison with benchmark schemes from the literature. 

\end{abstract}


\section{Introduction}
The ever-growing demand for higher data rates observed over the last few decades has become the main
challenge and research focus for the design of the next generation of wireless communication systems \cite{FSO_Survey_Demers}. In particular, it is expected that  by 2020 the number of devices which will use the fifth generation (5G) of wireless communication technology will reach tens or even hundreds of billions \cite{Corson} and the total required data rate will exceed 500 exabytes \cite{Hilton}.  Free space optical (FSO) systems are considered to be a powerful complementary and/or alternative  technology to the current radio frequency (RF) systems for meeting the data rate requirements of next generation wireless networks \cite{FSO_Survey_Demers}. In addition to the huge usable bandwidth, FSO systems are inherently secure and energy efficient \cite{FSO_Survey_Murat}.

The aforementioned beneficial properties of FSO systems come at the expense of some drawbacks and challenges which include the requirement of having a line of sight (LOS) between transmitter and receiver, the adverse effects of atmospheric turbulence, and unpredictable connectivity and temporary link outages due to visibility limiting conditions including snow, fog, and dust \cite{FSO_Survey_Murat, FSO_Survey_arXiv}. Various approaches have been proposed to mitigate these problems. For example, relay-based cooperation has been proposed as an effective strategy to facilitate an LOS between transmitter and receiver \cite{Relay_Selection_Chadi, Optimal_Relay_Placement_Murat}. Thereby, the parallel relaying network, where multiple relay nodes assist transmission from a source node to a destination node, is of particular interest \cite{Optimal_Relay_Placement_Murat, Relay_Selection_Chadi, FSO_Diversity_Chadi, Relay_Selection_Schober, GeorgeFSORelSelec}.  This network architecture provides spatial diversity which can be exploited to mitigate the fading induced by atmospheric turbulence. Moreover, since RF systems  are more reliable in terms of preserving connectivity albeit at lower data rates, hybrid RF/FSO systems, where an additional RF link is employed to support the FSO link, have been proposed. These systems can benefit from both the high data rates of the FSO link and the reliability of the RF link \cite{Alouini, Schober}.

The parallel FSO relay channel without RF backup links was considered in \cite{Optimal_Relay_Placement_Murat, Relay_Selection_Chadi, FSO_Diversity_Chadi, Relay_Selection_Schober}
 and  the parallel mixed RF/FSO relay channel with source-relay RF links and relay-destination FSO links was studied in \cite{GeorgeFSORelSelec}. Furthermore, the mixed RF/FSO relay channel with source-relay RF links and relay-destination hybrid RF/FSO links was considered in \cite{FSO_Vahid}.  However,  to  the  best  of  the authors' knowledge, the parallel hybrid RF/FSO relay channel, which is considered in this paper and its conference version  \cite{ICC_2016}, has not been investigated in the literature, yet. Such a communication system can be used for example for the wireless backhauling of a small-cell base station (BS) to a macro-cell BS \cite{Backhaul_VJ} and for forwarding data gathered by a wireless video surveillance camera to a central processing unit \cite{FSO_Survey_Murat} via multiple relays. Thereby, 
the nodes may be located on the roofs of buildings to maintain an LOS as required for FSO. The RF links support the FSO links in case of temporary loss of the LOS due to  adverse weather conditions or moving clouds and birds. We consider relay selection since it efficiently exploits the diversity that independent fading realizations offer and entails a significantly lower system complexity compared to transmission schemes where all  relays  are  active simultaneously \cite{GeorgeFSORelSelec,Relay_Selection_Schober,ParallelRelayAllFSO}. Furthermore, we assume full-duplex transmission for the FSO links owing to the narrow-beam property of FSO, whereas due to the broadcast nature of RF, half-duplex transmission is assumed for the RF links for the sake of simplicity and feasibility\footnote{Full-duplex RF relays have been reported in the literature \cite{DuarteFD}. However, they entail high hardware complexity for efficient self-interference suppression. Hence, in this paper, we focus on half-duplex RF relaying.}. 
For the relays, we consider two  cases depending on whether  or not they are equipped with buffers. For the non-buffer-aided (non-BA) case, the relay nodes receive data from the source and immediately forward it to the destination. On the other hand, for the buffer-aided (BA) case, the relay nodes can store the data received from the source in their buffers and forward it to the destination when their transmit channel qualities are  favorable \cite{Survay_BA}.

For both the non-BA and the BA cases, we derive the optimal relay selection policies for  the RF and FSO links such that the end-to-end  throughput is maximized. 
To further improve the throughput, the  time allocation between  RF transmission and reception for the selected relays is optimized. The proposed protocols provide important insights regarding optimal system design. For instance, the optimal non-BA policy selects at most two different relays for reception and transmission of the RF and FSO signals. In contrast, the optimal BA policy selects at most three different relays. Moreover,   we show that depending on which relays are selected for RF and FSO reception/transmission, there are three and ten possible optimal protocol modes for the non-BA and BA policies, respectively. These protocol modes can be further categorized into three types of transmission modes, namely the hybrid mode, the independent mode, and the mixed mode.  
We show that buffering can considerably enhance the throughput of the considered system at the expense of an increased  end-to-end delay \cite{Max_Max, Max_link}. Therefore, we also propose a delay-constrained BA policy which guarantees a certain target average delay. In addition, we develop distribution implementations for the optimal non-BA and BA policies. Our simulation results reveal that a considerable gain can be achieved  by the proposed optimal protocols in comparison with  benchmark schemes from the literature. Moreover, we show that the proposed delay-constrained BA protocol can approach the performance of the optimal delay-unconstrained BA protocol even for small average delays. 

We note that this paper is an extension of our conference paper \cite{ICC_2016} where only the non-BA case was studied. Moreover, this paper provides distributed implementations for the optimal policies, additional extensive discussions, simulation results, and rigorous proofs which are not included in \cite{ICC_2016}.  

The remainder of this paper is organized as follows. In Section~II, some preliminaries and assumptions are presented. In Section~III, the throughput maximization problems for both  the non-BA and the BA cases are formulated and the resulting optimal policies are derived. In Section~IV, we present solutions to two practical challenges of the proposed optimal policies, namely a delay-constrained BA protocol and distributed implementations for the optimal protocols. Simulation results are provided in Section~V and  conclusions are drawn in Section~VI.
 
\textit{Notations:} We use the following notations throughout this paper:  $\mathbbmss{E}\{\cdot\}$ denotes expectation,  $|\cdot|$ represents the magnitude of a complex number,  $\mathrm{erf}(\cdot)$ is the Gauss-error function, and $\mathrm{Pr}\{A\}$ denotes the probability of the occurrence of event $A$. Moreover, $\mathbf{0}$ denotes a vector with all elements equal to zero. Additionally, $\mathrm{Rice}(\Omega,\Psi)$ and $\mathrm{GGamma}(\Theta,\Phi)$ denote a Rician random variable (RV) with parameters $\Omega$ and $\Psi$ and a Gamma-Gamma RV with parameters $\Theta$ and $\Phi$, respectively.  For notational convenience, we use the definitions $\left[x\right]_a^b \triangleq \min\{b, \max \{a,x\} \}$ for $a\leq b$ and $\left[x\right]^+ \triangleq  \max \{0,x\} $.

\section{Preliminaries and Assumptions}\label{SysMod}
In this section, we present the considered system model, the channel models for the RF and FSO communication links, and the assumptions regarding the required channel state information (CSI).

\subsection{System Model}
The system model under consideration is schematically depicted in Fig.~\ref{FigSysMod} a). In particular, source $\mathcal{S}$ wishes to send its information to destination $\mathcal{D}$ via $M$ intermediate relay nodes denoted by $\mathcal{R}_m, \,\, m\in\lbrace1,\dots,M\rbrace $.  We assume that there is no direct link between $\mathcal{S}$ and $\mathcal{D}$. Moreover, the ${\mathcal{S}}-{\mathcal{R}}_m$ and ${\mathcal{R}}_{m}-{\mathcal{D}}$  links are hybrid RF/FSO where each FSO link is supported by an RF link. Fig.~\ref{FigSysMod} b) shows a possible application scenario of the considered communication system, namely wireless backhauling of a small-cell BS to a macro-cell BS via intermediate relays.  The entire time of operation is divided into $B$ equal-length  slots satisfying $B\to \infty$. Moreover, depending on whether or not the relay nodes are equipped with buffers, we consider two different cases namely  BA and non-BA relaying. Non-BA relay nodes ${\mathcal{R}}_{m}$ have to forward the data received from  $\mathcal{S}$ in the \textit{same time slot} to $\mathcal{D}$. In contrast, BA relays ${\mathcal{R}}_{m}$ are allowed to receive data from $\mathcal{S}$, store it in their buffers, and forward it to $\mathcal{D}$ when the ${\mathcal{R}}_{m}-{\mathcal{D}}$ link quality is favorable.

\iftoggle{OneColumn}{%

\begin{figure}
\centering
\scalebox{0.35}{
\pstool[width=2.5\linewidth]{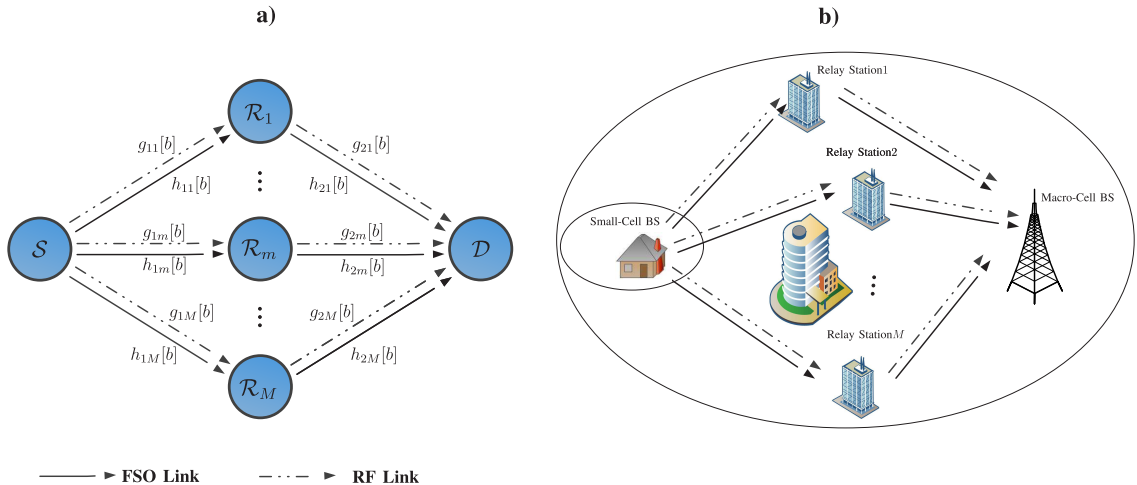}{
\psfrag{A}[c][c][2]{$\mathcal{S}$}
\psfrag{B}[c][c][2]{$\mathcal{D}$}
\psfrag{a}[c][c][2]{$\mathcal{R}_1$}
\psfrag{b}[c][c][2]{$\mathcal{R}_m$}
\psfrag{c}[c][c][2]{$\mathcal{R}_M$}
\psfrag{RF}[c][r][1.5]{$\textbf{RF Link}$}
\psfrag{FS}[c][r][1.5]{$\textbf{FSO Link}$}
\psfrag{gs1}[c][c][1.5]{$g_{11}[b]$}
\psfrag{gsm}[c][c][1.5]{$g_{1m}[b]$}
\psfrag{gsM}[c][c][1.5]{$g_{1M}[b]$}
\psfrag{hs1}[c][c][1.5]{$h_{11}[b]$}
\psfrag{hsm}[c][c][1.5]{$h_{1m}[b]$}
\psfrag{hsM}[c][c][1.5]{$h_{1M}[b]$}
\psfrag{g1d}[c][c][1.5]{$g_{21}[b]$}
\psfrag{gmd}[c][c][1.5]{$g_{2m}[b]$}
\psfrag{gMd}[c][c][1.5]{$g_{2M}[b]$}
\psfrag{h1d}[c][c][1.5]{$h_{21}[b]$}
\psfrag{hmd}[c][c][1.5]{$h_{2m}[b]$}
\psfrag{hMd}[c][c][1.5]{$h_{2M}[b]$}
\psfrag{R1}[c][c][2]{\textbf{a)}}
\psfrag{R2}[c][c][2]{\textbf{b)}}
\psfrag{SBS}[c][c][1]{Small-Cell BS}
\psfrag{RBS1}[c][c][1]{Relay Station $1$}
\psfrag{RBS2}[c][c][1]{Relay Station $2$}
\psfrag{RBSM}[c][c][1]{Relay Station $M$}
\psfrag{MBS}[c][c][1]{Macro-Cell BS}
}}
\caption{Parallel hybrid RF/FSO relay channel: a) schematic presentation and b) application scenario for wireless backhauling.}
\label{FigSysMod}\vspace{-0.3cm}
\end{figure}

}{%

\begin{figure*}
\centering
\scalebox{0.35}{
\pstool[width=2.5\linewidth]{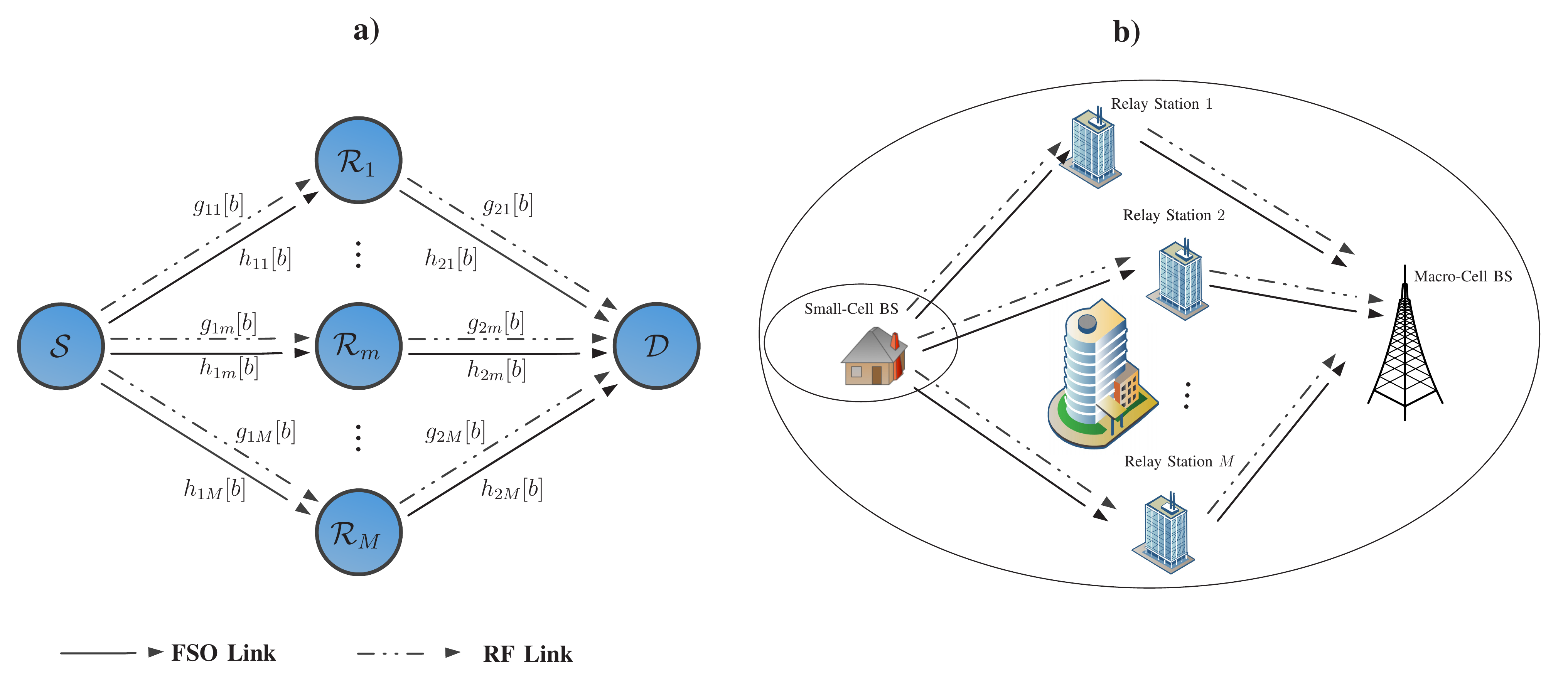}{
\psfrag{A}[c][c][2]{$\mathcal{S}$}
\psfrag{B}[c][c][2]{$\mathcal{D}$}
\psfrag{a}[c][c][2]{$\mathcal{R}_1$}
\psfrag{b}[c][c][2]{$\mathcal{R}_m$}
\psfrag{c}[c][c][2]{$\mathcal{R}_M$}
\psfrag{RF}[c][r][1.5]{$\textbf{RF Link}$}
\psfrag{FS}[c][r][1.5]{$\textbf{FSO Link}$}
\psfrag{gs1}[c][c][1.5]{$g_{11}[b]$}
\psfrag{gsm}[c][c][1.5]{$g_{1m}[b]$}
\psfrag{gsM}[c][c][1.5]{$g_{1M}[b]$}
\psfrag{hs1}[c][c][1.5]{$h_{11}[b]$}
\psfrag{hsm}[c][c][1.5]{$h_{1m}[b]$}
\psfrag{hsM}[c][c][1.5]{$h_{1M}[b]$}
\psfrag{g1d}[c][c][1.5]{$g_{21}[b]$}
\psfrag{gmd}[c][c][1.5]{$g_{2m}[b]$}
\psfrag{gMd}[c][c][1.5]{$g_{2M}[b]$}
\psfrag{h1d}[c][c][1.5]{$h_{21}[b]$}
\psfrag{hmd}[c][c][1.5]{$h_{2m}[b]$}
\psfrag{hMd}[c][c][1.5]{$h_{2M}[b]$}
\psfrag{R1}[c][c][2]{\textbf{a)}}
\psfrag{R2}[c][c][2]{\textbf{b)}}
\psfrag{SBS}[c][c][1]{Small-Cell BS}
\psfrag{RBS1}[c][c][1]{Relay Station $1$}
\psfrag{RBS2}[c][c][1]{Relay Station $2$}
\psfrag{RBSM}[c][c][1]{Relay Station $M$}
\psfrag{MBS}[c][c][1]{Macro-Cell BS}
}}
\caption{Parallel hybrid RF/FSO relay channel: a) schematic presentation and b) application scenario for wireless backhauling.}
\label{FigSysMod}\vspace{-0.3cm}
\end{figure*}
}

\subsection{Communication Links}
In the following, we describe the adopted channel model for  the FSO and RF links.
\subsubsection{FSO Links}
We assume that the FSO system employs
on-off keying (OOK) with intensity-modulation and direct-detection (IM/DD). 
Here, $\mathcal{S}$ is equipped with a multi-aperture transmitter pointing in the directions of the
relays.  Each relay has an aperture directed towards $\mathcal{D}$ and a photodetector for detection of the optical signal received from $\mathcal{S}$. Furthermore, $\mathcal{D}$ is equipped with a photodetector for detection of the optical signals received from the relays. Let $y_{1m}^{\mathrm{fso}}[b]$ and $y_{2m}^{\mathrm{fso}}[b]$ denote the intensities of the optical signals received at $\mathcal{R}_m$  and $\mathcal{D}$ in the $b$-th time slot, respectively. Thereby, after removing the ambient background light intensity, $y_{lm}^{\mathrm{fso}}[b]$ can be modelled as \cite{FSO_Survey_Murat, FSO_Survey_arXiv}
\begin{IEEEeqnarray}{lll}\label{Eq:Signal_FSO}
y_{lm}^{\mathrm{fso}}[b]=h_{lm}[b] x_{lm}^{\mathrm{fso}}[b]+z_{lm}^{\mathrm{fso}}[b],\,\,\,l=1,2,
\end{IEEEeqnarray}
where $x_{1m}^{\mathrm{fso}}[b]\in \lbrace 0,P^{\mathrm{fso}}_{\mathcal{S}} \rbrace$ and $x_{2m}^{\mathrm{fso}}[b]\in \lbrace 0,P^{\mathrm{fso}}_{\mathcal{R}_m} \rbrace$ denote the intensities of the  optical signals transmitted by $\mathcal{S}$ and  $\mathcal{R}_m$ in the $b$-th time slot, respectively. The maximum intensities of the FSO signals, i.e., $P^{\mathrm{fso}}_{\mathcal{S}}$ and $P^{\mathrm{fso}}_{\mathcal{R}_m}$, are mainly limited by restrictions  imposed by eye safety regulations \cite{FSO_Survey_Murat}. Moreover, $z_{1m}^{\mathrm{fso}}[b]$ and $z_{2m}^{\mathrm{fso}}[b]$ are the intensities of the shot noises caused by ambient light at $\mathcal{R}_m$ and  $\mathcal{D}$ in the $b$-th time slot, respectively. Noises $z_{1m}^{\mathrm{fso}}[b]$ and $z_{2m}^{\mathrm{fso}}[b]$ are modelled as zero-mean real-valued additive white Gaussian noises (AWGNs) with variances $\sigma_{1m}^2$ and $\sigma_{2m}^2$, respectively, and are independent from each other and from the  transmitted FSO signals. Furthermore, $h_{1m}[b]$ and $h_{2m}[b]$ denote the channel gains of the ${\mathcal{S}}-{\mathcal{R}}_m$ and ${\mathcal{R}}_{m}-{\mathcal{D}}$ FSO links in the $b$-th time slot, respectively, and are modelled as mutually independent, ergodic, and stationary random processes with continuous probability density functions (pdfs).  We adopt the widely-accepted Gamma-Gamma  turbulence model \cite{FSO_Survey_Murat, HybridRFFSOShi, Schober}. Hence, $h_{lm}[b],\,\,l=1,2,$ is modelled as $h_{lm}[b]=\bar{h}_{lm}\tilde{h}_{lm}[b]$, where $\bar{h}_{lm}$ and $\tilde{h}_{lm}[b]$ are the average gain and the fading gain of the FSO links, respectively, and are given by \cite{FSO_Survey_Murat, HybridRFFSOShi, Schober}
\begin{IEEEeqnarray}{rll}\label{Eq:FSO_LinkModel}
\begin{cases}
\bar{h}_{lm}=\mathit{R}\left[ \mathrm{erf}\left( \dfrac{\sqrt{\pi}r}{\sqrt{2}\phi d_{lm}}\right) \right] ^2\times 10^{-k_{lm}d_{lm}/10}\\
\tilde{h}_{lm}[b] \sim \mathrm{GGamma}(\Theta,\Phi),
\end{cases}
\end{IEEEeqnarray}
where $\mathit{R}$ denotes the responsivity of the photodetector, $r$ is the aperture radius, $\phi$ is the divergence angle of the beam, $d_{1m}$ and $d_{2m}$ are the distances between the transmitters and the receivers of the $\mathcal{S}-\mathcal{R}_m$ and $\mathcal{R}_m-\mathcal{D}$ links, respectively, and $k_{1m}$ and $k_{2m}$ are the weather-dependent attenuation factors of the $\mathcal{S}-\mathcal{R}_m$ and $\mathcal{R}_m-\mathcal{D}$ FSO links, respectively. Parameters $\Theta$ and $\Phi$ of the Gamma-Gamma distribution depend on physical parameters such as the wavelength $\lambda^{\mathrm{fso}}$ and the weather-dependent index of refraction structure parameter $C_n^2$, cf. \cite[Eqs. (3) and (4)]{Schober}. 

From an information theoretical point of view, the considered FSO links can be modelled as binary input-continuous output AWGN channels where the maximum information rate is achieved by uniformly distributed  binary inputs \cite{OOKCapacity}.  In the $b$-th time slot, the capacities of the $\mathcal{S}-\mathcal{R}_m$ and $\mathcal{R}_m-\mathcal{D}$  FSO links, denoted by $C_{1m}^{\mathrm{fso}}[b]$ and $C_{2m}^{\mathrm{fso}}[b]$, respectively, for OOK inputs are given by \cite{OOKCapacity}
\iftoggle{OneColumn}{%
\begin{IEEEeqnarray}{lll}\label{Eq:FSO_C}
C_{lm}^{\mathrm{fso}}[b]=&W^{\mathrm{fso}}\Bigg[1-\dfrac{1}{\sqrt{2\pi}}\displaystyle\int\limits_{ - \infty }^\infty\mathrm{exp} (-t^2)  \mathrm{log}_2 \Bigg\lbrace 1+\mathrm{exp}\left(  - \dfrac{p_{lm}^2[b]}{2\sigma_{lm}^2}\right) \nonumber\\
&\left[ \mathrm{exp}\left(  \dfrac{2tp_{lm}[b]}{\sqrt{2\sigma_{lm}^2}}\right) +\mathrm{exp}\left(  - \dfrac{2tp_{lm}[b]}{\sqrt{2\sigma_{lm}^2}}\right) +\mathrm{exp}\left(  - \dfrac{p_{lm}^2[b]}{2\sigma_{lm}^2}\right) \right]  \Bigg\rbrace \mathrm{d}t\Bigg],\,\,
\end{IEEEeqnarray}
}{%
\begin{IEEEeqnarray}{lll}\label{Eq:FSO_C}
C_{lm}^{\mathrm{fso}}[b]=&W^{\mathrm{fso}}\Bigg[1-\dfrac{1}{\sqrt{2\pi}}\displaystyle\int\limits_{ - \infty }^\infty\mathrm{exp} (-t^2)  \\
 & \mathrm{log}_2 \Bigg\lbrace 1+\mathrm{exp}\left(  - \dfrac{p_{lm}^2[b]}{2\sigma_{lm}^2}\right) \Bigg[ \mathrm{exp}\left(  \dfrac{2tp_{lm}[b]}{\sqrt{2\sigma_{lm}^2}}\right)
 \nonumber\\
& +\mathrm{exp}\left(  - \dfrac{2tp_{lm}[b]}{\sqrt{2\sigma_{lm}^2}}\right) +\mathrm{exp}\left(  - \dfrac{p_{lm}^2[b]}{2\sigma_{lm}^2}\right) \Bigg]  \Bigg\rbrace \mathrm{d}t\Bigg],\,\, \nonumber
\end{IEEEeqnarray}
}
where $p_{1m}[b]=P^{\mathrm{fso}}_\mathcal{S}h_{1m}[b]$, $p_{2m}[b]=P^{\mathrm{fso}}_{\mathcal{R}_m}h_{2m}[b]$, and $W^{\mathrm{fso}}$ is the bandwidth of the FSO signal.  
\subsubsection{RF Links}
We consider a standard AWGN channel for the RF links. Moreover, we assume that all RF transmitters and receivers are equipped with a single antenna. Let $y_{1m}^{\mathrm{rf}}[b]$ and $y_{2m}^{\mathrm{rf}}[b]$ denote the RF signals received at $\mathcal{R}_m$ and $\mathcal{D}$ in the $b$-th time slot, respectively, and be modelled as \cite{Cover}
\begin{IEEEeqnarray}{lll}\label{Eq:Signal_RF}
y_{lm}^{\mathrm{rf}}[b]=g_{lm}[b]x_{lm}^{\mathrm{rf}}[b]+z_{lm}^{\mathrm{rf}}[b],\,\,\,l=1,2,
\end{IEEEeqnarray}
where $x_{1m}^{\mathrm{rf}}[b]$ and $x_{2m}^{\mathrm{rf}}[b]$ are the  RF signals transmitted by $\mathcal{S}$  and   $\mathcal{R}_m$, respectively. Additionally, $z_{1m}^{\mathrm{rf}}[b]$ and $z_{2m}^{\mathrm{rf}}[b]$ denote the receiver noises at  $\mathcal{R}_m$ and  $\mathcal{D}$ in the $b$-th time slot, respectively. We assume that $z_{1m}^{\mathrm{rf}}[b]$ and $z_{2m}^{\mathrm{rf}}[b]$ can be modelled as zero-mean complex AWGNs with variances $\delta_{1m}^2$ and $\delta_{2m}^2$, respectively. The RF noise variances are given by $[\delta_{lm}^2]_{\mathrm{dB}}=W^\mathrm{rf}N_{lm,0}+N_{lm,F}$, where $W^\mathrm{rf}$ is the bandwidth of the RF signal, $N_{lm,0}$ denotes the noise power spectral density (in dB/Hz), and $N_{lm,F}$ is the noise figure (in dB) of the RF receivers. Furthermore, $g_{1m}[b]$ and $g_{2m}[b]$ are mutually independent, ergodic, and stationary random processes with continuous pdfs specifying the channel coefficients of the ${\mathcal{S}}-{\mathcal{R}}_m$ and ${\mathcal{R}}_{m}-{\mathcal{D}}$ RF links in the $b$-th time slot, respectively. For the hybrid RF/FSO link, an LOS has to be available  for  the applicability of the FSO system \cite{HybridRFFSOShi, Schober}. Therefore, we assume Rician fading for the RF links which includes the effects of both scattered and LOS paths. Taking into account the effect of path-loss,  $g_{lm}[b]$ is modelled as $g_{lm}[b]=\sqrt{\bar{g}_{lm}}\tilde{g}_{lm}[b]$, where $\bar{g}_{lm}$ and $\tilde{g}_{lm}[b]$ denote the average gain and the fading coefficient of the RF links, respectively, and are given by \cite{WiMAXFSO,Schober}
\begin{IEEEeqnarray}{rll}\label{Eq:RF_LinkModel}
\begin{cases}
\bar{g}_{lm}=\Bigg[\dfrac{\lambda^{\mathrm{rf}}\sqrt{G_{\mathrm{tx}}^{\mathrm{rf}}G_{\mathrm{rx}}^{\mathrm{rf}}}}{4 \pi d^{\mathrm{rf}}_{\mathrm{ref}}}\Bigg]^2\times \Bigg[\dfrac{d^{\mathrm{rf}}_{\mathrm{ref}}}{d_{lm}}\Bigg]^{\nu_{lm}}\\
|\tilde{g}_{lm}[b]|  \sim \mathrm{Rice}(\Omega,\Psi),
\end{cases}
\end{IEEEeqnarray}
where $\lambda^{\mathrm{rf}}$ is the wavelength of the RF signal, $G_{\mathrm{tx}}^{\mathrm{rf}}$ and $G_{\mathrm{rx}}^{\mathrm{rf}}$ are the transmit and receive RF antenna gains, respectively, and $d^{\mathrm{rf}}_{\mathrm{ref}}$ denotes a reference distance for the antenna far-field. Moreover, $\nu_{1m}$ and $\nu_{2m}$ are the path-loss exponents of the $\mathcal{S}-\mathcal{R}_m$ and $\mathcal{R}_m-\mathcal{D}$ RF links, respectively. Parameters $\Omega$ and $\Psi$ of the Rice distribution denote the ratios between the power in the direct path and the power in the scattered paths to the total power in both paths, respectively. Moreover, the capacities of the $\mathcal{S}-\mathcal{R}_m$ and $\mathcal{R}_m-\mathcal{D}$ RF links in the $b$-th time slot, denoted by $C_{1m}^{\mathrm{rf}}[b]$ and $C_{2m}^{\mathrm{rf}}[b]$, respectively,  are given by
\begin{IEEEeqnarray}{lll}\label{Eq:RF_C}
C_{lm}^{\mathrm{rf}}[b]=W^{\mathrm{rf}}\mathrm{log_2}\left(1+\dfrac{ q_{lm}^2[b]}{\delta^2_{lm}}\right),\,\,l=1,2,
\end{IEEEeqnarray}
where $q_{1m}[b]=\sqrt{P^{\mathrm{rf}}_{\mathcal{S}}}|g_{1m}[b]|$ and $q_{2m}[b]=\sqrt{P^{\mathrm{rf}}_{\mathcal{R}_m}}|g_{2m}[b]|$. Here, $P^{\mathrm{rf}}_{\mathcal{S}}$ and $P^{\mathrm{rf}}_{\mathcal{R}_m}$ are the RF transmit powers of $\mathcal{S}$ and $\mathcal{R}_m$, respectively.

\begin{remk}
In this paper, we assume OOK signaling for the FSO links and Gaussian signaling for the RF links. However, we note that the considered problem formulation and the resulting non-BA and BA policies given in the next section are given in general form such that they are also applicable if different signaling schemes are adopted for the RF and FSO links. In particular, for other signaling schemes, only the expressions in (\ref{Eq:FSO_C}) and (\ref{Eq:RF_C}) have to be modified and then be used in the proposed relay selection policies presented in Section III. 
\end{remk}

\subsection{CSI Requirements}

In Section~III, we derive the optimal non-BA and BA policies assuming that a central node, e.g., the destination, has the instantaneous CSI of all FSO and RF links and  is responsible for determining the transmission strategy and conveying it to all other nodes. However, in Subsection IV.B, we present distributed implementations of the optimal policies where each node needs to acquire only the CSI of those RF/FSO links to which it is directly connected. Typically, in hybrid RF/FSO systems, the coherence time of the RF links is on the order of seconds whereas the coherence time of the FSO links is on the order of milliseconds \cite{Coherenece}. Therefore, for time slot durations on the order of milliseconds, the hybrid RF/FSO channel is constant and can accommodate thousands of RF/FSO symbols per time slot for typical RF/FSO symbol rates.
Because of the large coherence time, we assume that the signaling overhead caused by channel estimation and feedback is negligible compared to the amount of information transmitted in one time slot.

\section{Throughput-Optimal Relay Selection Policies}

In this section, we first present the problem formulation for relay selection, and subsequently, we derive the optimal non-BA and  BA  policies maximizing  the throughput as functions of the fading state.

\subsection{Problem Formulation for Relay Selection}

 For the considered communication system, our goal is to derive  optimal relay selection policies which maximize the throughput for both non-BA and BA relays given the CSI of all RF and FSO links. To this end, let $\alpha_{1m}[b],\,\, m\in\lbrace1,\dots,M\rbrace$, denote binary selection variables where $\alpha_{1m}[b]=1$ if relay $\mathcal{R}_
 m$ is selected for FSO reception in the $b$-th time slot and $\alpha_{1m}[b]=0$ if relay $\mathcal{R}_m$ is not selected. Similarly,  $\alpha_{2m}[b]=1$ indicates that relay $\mathcal{R}_m$ is selected for FSO transmission in the $\mathit{b}$-th time slot and $\alpha_{2m}[b]=0$ if relay $\mathcal{R}_m$ is not selected. Analogously, $\beta_{1m}[\mathit{b}]$ and $ \beta_{2m}[b],\,\,m\in\lbrace1,\dots,M\rbrace$, denote binary selection variables  for RF relay selection for reception and transmission in the $b$-th time slot, respectively. For simplicity of implementation, we assume that in each time slot,  one relay is selected for RF reception and one relay is selected for FSO reception. Similarly,  one relay is selected for RF transmission and one relay is selected for FSO transmission. We note that  activation of multiple relay nodes for simultaneous reception or transmission requires more complicated transmission schemes because of the required multi-user encoding/decoding. In addition, it is known that in general, despite its simplicity, relay selection efficiently exploits the diversity gain that independent fading realizations provide \cite{GeorgeFSORelSelec,Relay_Selection_Schober,ParallelRelayAllFSO}.  Mathematically, in order to enforce the aforementioned assumptions on the relay selection strategy, 
 $\sum_{\forall m}\alpha_{lm}[b]=1,\,\,\forall l,b$, and $\sum_{\forall m}\beta_{lm}[b]=1,\,\,\forall l,b$, have to hold.  

 Due to the broadcast nature of RF, simultaneous activation of the selected relays creates interference from the transmitting relay to the receiving relay. In particular,  self-interference occurs if the same relay is selected for both RF transmission and RF reception and inter-relay interference occurs if the relays selected  for RF transmission and RF reception are different. Therefore, for the sake of simplicity of implementation and practical feasibility, we assume that the RF links are half duplex with respect to each other. In other words, assuming relays $\mathcal{R}_n$ and  $\mathcal{R}_{n'}$ are selected for RF reception and RF transmission, respectively, the ${\mathcal{S}}-{\mathcal{R}}_n$ and ${\mathcal{R}}_{n'}-{\mathcal{D}}$ RF links cannot be active at the same time. Hence, we activate the ${{\mathcal{S}}-{\mathcal{R}}_n}$ RF link in the $\rho_{1}[b]\in[0,1]$ fraction of the $b$-th time slot and the ${{\mathcal{R}}_{n'}-{\mathcal{D}}}$ RF link in the remaining  $\rho_2[b]\in[0,1]$ fraction of the $b$-th time slot, respectively, where $\rho_{1}[b]+\rho_{2}[b]=1,\,\,\forall b$, holds.
 
  On the other hand, assuming relays $\mathcal{R}_m$ and  $\mathcal{R}_{m'}$ are selected for FSO reception and transmission, respectively, they can simultaneously transmit over both the ${\mathcal{S}}-{\mathcal{R}}_m$ and ${\mathcal{R}}_{m'}-{\mathcal{D}}$ FSO links, i.e., the FSO links are orthogonal with respect to each other due to narrow-beam property of FSO. In the BA case, the relays can extract data from their buffers and send it to the destination at the same time when they are receiving data from the source. This allows the source and the relays to construct codewords which span one time slot. However, in the non-BA case, if the source codeword spans one time slot, the relays have to wait until the end of the time slot before they can decode the FSO signal. Therefore, the relays cannot forward this data to the destination in the same time slot which contradicts the basic assumption behind non-BA transmission, namely that the data transmitted by the source has to be received by the destination in the same time slot. To alleviate this problem, we assume that for non-BA transmission, each time slot is divided into $n$ sub-slots indexed by $i=1,2,\dots,n$. Thereby, the relays can transmit the  data received from the source in sub-slot $i=1,2,\dots,n-1$ to the destination in the subsequent sub-slot $i+1$. Thereby, the effective capacities of the  FSO links is $\dfrac{n-1}{n}C_{lm}^{\mathrm{fso}}[b]$ which approaches $C_{lm}^{\mathrm{fso}}[b]$ as $n \to \infty$, i.e., the full-duplex property of the FSO links is fully exploited. The considered transmission protocol is schematically illustrated in Fig.~2.

\iftoggle{OneColumn}{%
\begin{figure}
\centering
\scalebox{0.6}{
\pstool[width=1\linewidth]{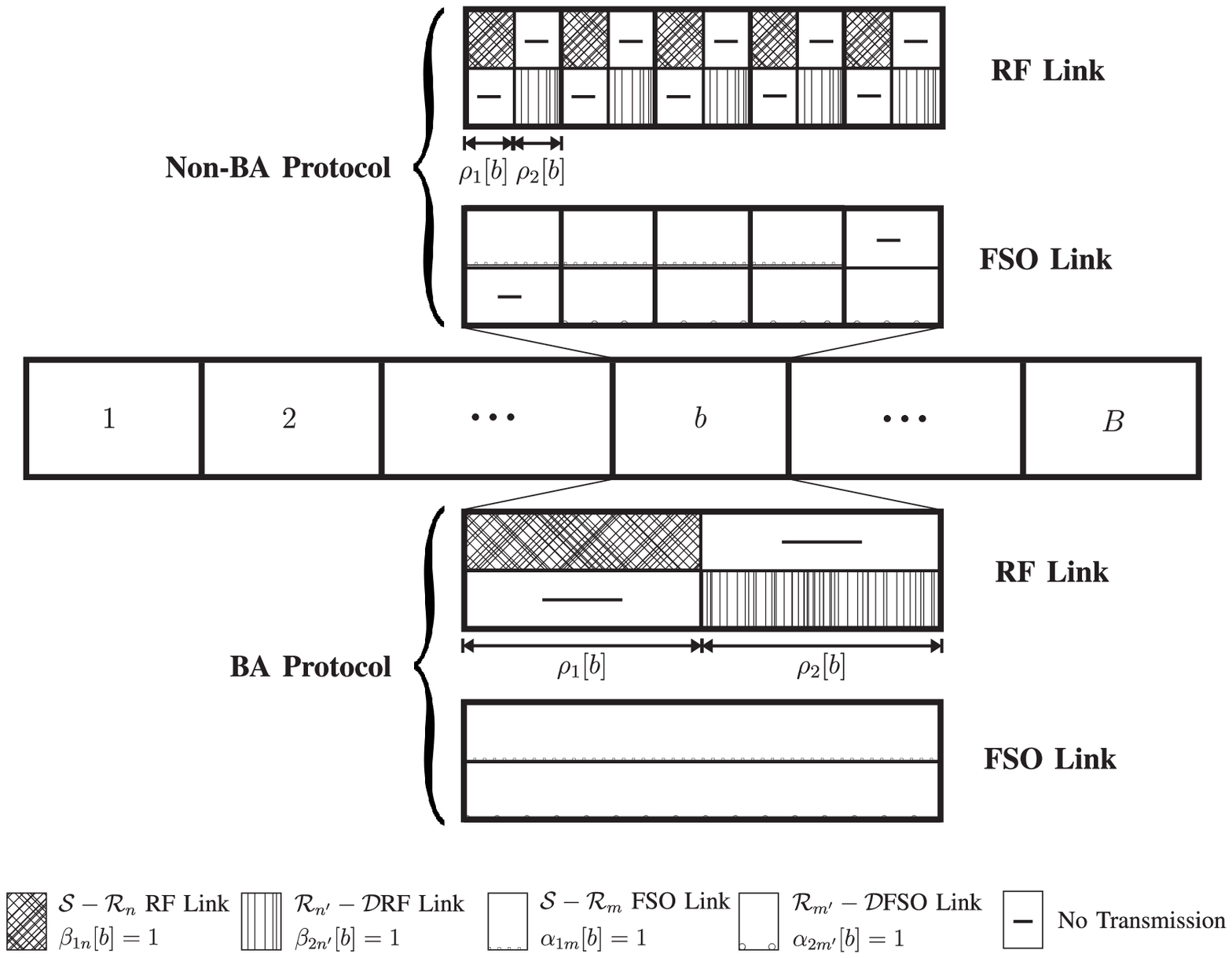}{
\psfrag{1}[c][c][1]{$1$}
\psfrag{2}[c][c][1]{$2$}
\psfrag{b}[c][c][1]{$b$}
\psfrag{B}[c][c][1]{$B$}
\psfrag{RF}[c][r][1]{$\textbf{RF Link}$}
\psfrag{FS}[c][r][1]{$\textbf{FSO Link}$}
\psfrag{NB}[r][r][1]{$\textbf{Non-BA Protocol}$}
\psfrag{BA}[r][r][1]{$\textbf{BA Protocol}$}
\psfrag{g1}[l][c][0.8]{${{\mathcal{S}}-{\mathcal{R}}_n}$ RF Link}
\psfrag{g2}[l][c][0.8]{${{\mathcal{R}}_{n'}-{\mathcal{D}}}$ RF Link}
\psfrag{h1}[l][c][0.8]{${{\mathcal{S}}-{\mathcal{R}}_{m}}$ FSO Link}
\psfrag{h2}[l][c][0.8]{${{\mathcal{R}}_{m'}-{\mathcal{D}}}$ FSO Link}
\psfrag{t1}[c][c][0.9]{$\rho_1[b]$}
\psfrag{t2}[c][c][0.9]{$\rho_2[b]$}
\psfrag{c1}[l][c][0.8]{$\beta_{1n}[b]=1$}
\psfrag{c2}[l][c][0.8]{$\beta_{2n'}[b]=1$}
\psfrag{a1}[l][c][0.8]{$\alpha_{1m}[b]=1$}
\psfrag{a2}[l][c][0.8]{$\alpha_{2m'}[b]=1$}
\psfrag{n1}[l][c][0.8]{No Transmission}
}}
\caption{Proposed transmission protocol for the considered parallel hybrid RF/FSO relay channel.\vspace{-0.3cm}}
\label{FigMod}
\end{figure}
}{%
\begin{figure*}
\centering
\scalebox{0.6}{
\pstool[width=1\linewidth]{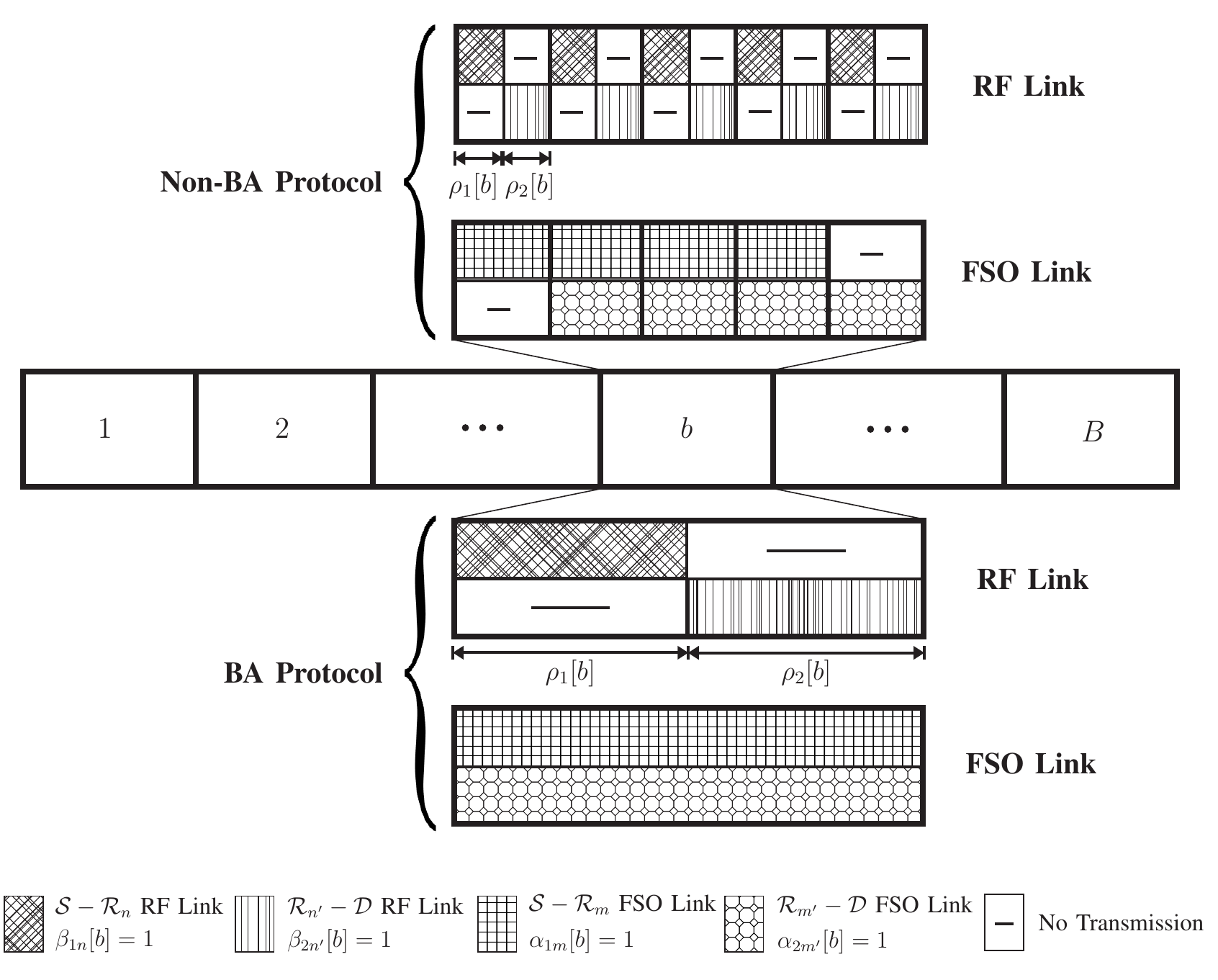}{
\psfrag{1}[c][c][1]{$1$}
\psfrag{2}[c][c][1]{$2$}
\psfrag{b}[c][c][1]{$b$}
\psfrag{B}[c][c][1]{$B$}
\psfrag{RF}[c][r][1]{$\textbf{RF Link}$}
\psfrag{FS}[c][r][1]{$\textbf{FSO Link}$}
\psfrag{NB}[r][r][1]{$\textbf{Non-BA Protocol}$}
\psfrag{BA}[r][r][1]{$\textbf{BA Protocol}$}
\psfrag{g1}[l][c][0.8]{${{\mathcal{S}}-{\mathcal{R}}_n}$ RF Link}
\psfrag{g2}[l][c][0.8]{${{\mathcal{R}}_{n'}-{\mathcal{D}}}$ RF Link}
\psfrag{h1}[l][c][0.8]{${{\mathcal{S}}-{\mathcal{R}}_{m}}$ FSO Link}
\psfrag{h2}[l][c][0.8]{${{\mathcal{R}}_{m'}-{\mathcal{D}}}$ FSO Link}
\psfrag{t1}[c][c][0.9]{$\rho_1[b]$}
\psfrag{t2}[c][c][0.9]{$\rho_2[b]$}
\psfrag{c1}[l][c][0.8]{$\beta_{1n}[b]=1$}
\psfrag{c2}[l][c][0.8]{$\beta_{2n'}[b]=1$}
\psfrag{a1}[l][c][0.8]{$\alpha_{1m}[b]=1$}
\psfrag{a2}[l][c][0.8]{$\alpha_{2m'}[b]=1$}
\psfrag{n1}[l][c][0.8]{No Transmission}
}}
\caption{Proposed transmission protocol for the considered parallel hybrid RF/FSO relay channel.\vspace{-0.3cm}}
\label{FigMod}
\end{figure*}
}

\subsection{Optimal Non-BA Policy}

In this subsection, we derive the optimal \textit{adaptive} non-BA RF/FSO  relay selection and RF transmission time allocation policies  such that the average information rate from the source to the destination, denoted by $\bar{\tau}$, is maximized. The resulting throughput maximization problem can be formulated as 
\iftoggle{OneColumn}{%
\begin{IEEEeqnarray}{cll}\label{Eq:Opt_DL}
\underset{\boldsymbol{\alpha}\in{{\boldsymbol{\mathcal{A}}}},\boldsymbol{\beta}\in{{\boldsymbol{\mathcal{B}}}},\boldsymbol{\rho}\in{{\boldsymbol{\mathcal{C}}}},\boldsymbol{\tau}\geq \mathbf{0}}{\mathrm{maximize}}\,\,&\bar{\tau}=\sum_{\forall m}\bar{\tau}_m =\dfrac{1}{B} \sum_{\forall b} \sum_{\forall m} \tau_m[b] \\
\mathrm{subject~to} \,\, & \tau_m[b] \leq \alpha_{1m}[b]  C_{1m}^{\mathrm{fso}}[b] +\beta_{1m}[b] \rho_1[b]  C_{1m}^{\mathrm{rf}}[b], \quad \forall m,b, \nonumber \\
&\tau_m[b] \leq \alpha_{2m}[b]  C_{2m}^{\mathrm{fso}}[b] +\beta_{2m}[b] \rho_2[b]  C_{2m}^{\mathrm{rf}}[b], \quad \forall m,b,\nonumber
\end{IEEEeqnarray}
}{%
\begin{IEEEeqnarray}{cll}\label{Eq:Opt_DL}
\underset{\boldsymbol{\alpha}\in{{\boldsymbol{\mathcal{A}}}},\boldsymbol{\beta}\in{{\boldsymbol{\mathcal{B}}}},\boldsymbol{\rho}\in{{\boldsymbol{\mathcal{C}}}},\boldsymbol{\tau}\geq \mathbf{0}}{\mathrm{maximize}}
\,\,&\bar{\tau}=\sum_{\forall m}\bar{\tau}_m =\dfrac{1}{B} \sum_{\forall b} \sum_{\forall m} \tau_m[b] \\
\mathrm{subject~to} \,\, 
& \tau_m[b] \leq \alpha_{1m}[b]  C_{1m}^{\mathrm{fso}}[b] \nonumber \\
& \quad\qquad +\beta_{1m}[b] \rho_1[b]  C_{1m}^{\mathrm{rf}}[b], \quad \forall m,b, \nonumber \\
& \tau_m[b] \leq \alpha_{2m}[b]  C_{2m}^{\mathrm{fso}}[b]  \nonumber \\
& \quad\qquad +\beta_{2m}[b] \rho_2[b]  C_{2m}^{\mathrm{rf}}[b], \quad \forall m,b,\nonumber
\end{IEEEeqnarray}
}
where $\boldsymbol{\alpha}$, $\boldsymbol{\beta}$, $\boldsymbol{\rho}$, and $\boldsymbol{\tau}$  are the vectors containing the relay selection variables of the FSO links, the relay selection variables of the RF links, the time sharing variables of the RF links, and the relays' throughputs, respectively. We note that since the optimal non-BA  policy depends only on the fading states of the FSO and RF links, and not on the transmission time slot index, we drop the time slot index in this subsection for notational simplicity.
Moreover, $\boldsymbol{\mathcal{A}}=\{\boldsymbol{\alpha}|\alpha_{lm}  \in \{0,1\},\,\, \forall l,m\,\,\wedge\,\,\sum_{\forall m}\alpha_{lm}=1,\,\,\forall l \}$, $\boldsymbol{\mathcal{B}}=\{\boldsymbol{\beta}|\beta_{lm} \in \{0,1\},\,\, \forall l,m\,\,\wedge\,\,\sum_{\forall m}\beta_{lm}=1,\,\,\forall l \}$, and $\boldsymbol{\mathcal{C}}=\{\boldsymbol{\rho}|\rho_{l} \in [0,1],\,\, \forall l\,\,\wedge\,\,\sum_{\forall l}\rho_{l}=1 \}$ are the feasible sets of $\boldsymbol{\alpha}$, $\boldsymbol{\beta}$, and ${\boldsymbol{\rho}}$, respectively. The constraints  in (\ref{Eq:Opt_DL}) follow from the max-flow min-cut theorem \cite{Cover}, according to which the throughput of relay $\mathcal{R}_m$ is limited by 
the capacities of the $\mathcal{S}-\mathcal{R}_m$ and $\mathcal{R}_m-\mathcal{D}$ links, respectively. In the following theorem, the optimal solution to the optimization problem in (\ref{Eq:Opt_DL}) is provided. 

\begin{theo}\label{Theo:Delay-Limited}
For the parallel non-BA relay channel with hybrid RF/FSO links, the optimal policies for FSO and RF relay selection and optimal RF transmission time allocation are given by
\iftoggle{OneColumn}{%
\begin{IEEEeqnarray}{rll}\label{Eq:DL-theo}
 \alpha_{lm^*}    
 &\Equal {\begin{cases}
1,  &\mathrm{if}\,\, \big\{ \text{Case 1} \,\,\wedge\,\, m^* = \underset{m}{\mathrm{argmax}}\,\, \tau_{m}^{\mathrm{hyb}} \big\}\\
&\vee \,\, \big\{ \text{Case 2} \,\,\wedge\,\,  m^*  = \underset{m}{\mathrm{argmax}}\,\, \tau_{m}^{\mathrm{fso}} \big\}\\
&\vee \,\, \big\{ \text{Case 3} \, \wedge\,  l = 1 \, \wedge\, (m^*,-) = \underset{(m,n)}{\mathrm{argmax}}\,\, \tau_{mn}^{\mathrm{mix}} \big\}\\
&\vee \,\, \big\{ \text{Case 3} \, \wedge\,  l = 2 \, \wedge\, (-,m^*) = \underset{(m,n)}{\mathrm{argmax}}\,\, \tau_{mn}^{\mathrm{mix}} \big\}\\
0, &\mathrm{otherwise}
\end{cases}}   \quad\,\,\, \IEEEyesnumber\IEEEyessubnumber 
\\
 \beta_{lm^*} 
&\Equal  {\begin{cases}
1,  &\mathrm{if}\,\, \big\{ \text{Case 1} \,\,\wedge\,\, m^* = \underset{m}{\mathrm{argmax}}\,\, \tau_{m}^{\mathrm{hyb}}  \big\}\\
&\vee \,\, \big\{ \text{Case 2} \,\,\wedge\,\, m^* = \underset{m}{\mathrm{argmax}}\,\, \tau_{m}^{\mathrm{rf}} \big\}\\
&\vee \,\, \big\{ \text{Case 3} \, \wedge\,  l = 1 \, \wedge\, (-,m^*) = \underset{(m,n)}{\mathrm{argmax}}\,\, \tau_{mn}^{\mathrm{mix}}  \big\}\\
&\vee \,\, \big\{ \text{Case 3} \, \wedge\,  l = 2 \, \wedge\, (m^*,-) = \underset{(m,n)}{\mathrm{argmax}}\,\, \tau_{mn}^{\mathrm{mix}}  \big\}\\
0, &\mathrm{otherwise}
\end{cases}}  \IEEEyessubnumber \\
 \rho_1^* \Equal 1 \Minus \rho_2^*    
&\Equal  \begin{cases}
\left[ \frac{C_{2m^*}^{\mathrm{fso}} +C_{2m^*}^{\mathrm{rf}} -C_{1m^*}^{\mathrm{fso}} }{C_{1m^*}^{\mathrm{rf}} +C_{2m^*}^{\mathrm{rf}} } \right]_0^1, 
&  \mathrm{if}\,\, \big\{ \text{Case 1} \,\,\wedge\,\, m^* = \underset{m}{\mathrm{argmax}}\,\, \tau_{m}^{\mathrm{hyb}}  \big\}\\
 \frac{C_{2m^*}^{\mathrm{rf}} }{C_{1m^*}^{\mathrm{rf}} +C_{2m^*}^{\mathrm{rf}} }, \,\,  &\mathrm{if}\,\, \big\{ \text{Case 2} \,\,\wedge\,\, m^* = \underset{m}{\mathrm{argmax}}\,\, \tau_{m}^{\mathrm{rf}}  \big\}\\
\frac{C_{2m^*}^{\mathrm{fso}} }{C_{1m^*}^{\mathrm{rf}} }, \qquad &\mathrm{if}\,\, \big\{ \text{Case 3} \,\,\wedge\,\, (-,m^*) = \underset{(m,n)}{\mathrm{argmax}}\,\, \tau_{mn}^{\mathrm{mix}}  \big\}\\
 \end{cases} \hspace{-0.3cm} \IEEEyessubnumber
\end{IEEEeqnarray}
where $\tau_{m}^{\mathrm{hyb}}$,  $\tau_{mn}^{\mathrm{ind}}$, and $\tau_{mn}^{\mathrm{mix}}$ are given by
\begin{IEEEeqnarray}{lll} \label{Eq:DiffThr}
\Scale[1]{ \tau_m^{\mathrm{hyb}}=\begin{cases}
C_{2m}^{\mathrm{fso}}+C_{2m}^{\mathrm{rf}}, \qquad \mathrm{if}\,\, \frac{C_{2m}^{\mathrm{fso}}+C_{2m}^{\mathrm{rf}}}{ C_{1m}^{\mathrm{fso}}}<1 \\
C_{1m}^{\mathrm{fso}}+C_{1m}^{\mathrm{rf}}, \qquad  \mathrm{if}\,\, \frac{C_{1m}^{\mathrm{fso}}+C_{1m}^{\mathrm{rf}}}{C_{2m}^{\mathrm{fso}}}<1 \\
\frac{C_{2m}^{\mathrm{fso}}+C_{2m}^{\mathrm{rf}}-C_{1m}^{\mathrm{fso}}}{C_{1m}^{\mathrm{rf}}+C_{2m}^{\mathrm{rf}}} C_{1m}^{\mathrm{rf}} + C_{1m}^{\mathrm{fso}},   \,\, \mathrm{otherwise}
\end{cases} }\IEEEyesnumber\IEEEyessubnumber \\
\Scale[1]{\tau_{mn}^{\mathrm{ind}}  =  \tau_{m}^{\mathrm{fso}}  + \tau_{n}^{\mathrm{rf}},\,\,\,  \mathrm{where} \,\,\,
\begin{cases} \tau_{m}^{\mathrm{fso}} = \min\big\{C_{1m}^{\mathrm{fso}},C_{2m}^{\mathrm{fso}}\big\} \\  \tau_{n}^{\mathrm{rf}} = \frac{C_{1n}^{\mathrm{rf}} C_{2n}^{\mathrm{rf}}}{C_{1n}^{\mathrm{rf}}+C_{2n}^{\mathrm{rf}}} \end{cases}}   \IEEEyessubnumber \\
\Scale[1]{\tau_{mn}^{\mathrm{mix}} =  \begin{cases}
C_{1m}^{\mathrm{fso}}+C_{2n}^{\mathrm{fso}},\,\,\,\,&\mathrm{if}\,\,\frac{C_{2n}^{\mathrm{fso}}}{C_{1n}^{\mathrm{rf}}}+\frac{C_{1m}^{\mathrm{fso}}}{C_{2m}^{\mathrm{rf}}}\leq 1 \\
0,&\mathrm{otherwise.}
\end{cases}}   \IEEEyessubnumber  
\end{IEEEeqnarray}
 Moreover, Cases 1-3 are defined as follows
\begin{IEEEeqnarray}{rll} \label{Eq:Cases}
\Scale[1]{
\begin{cases}
\text{Case 1 (Hybrid Mode):} \\
\quad \underset{m}{\mathrm{max}}\,\, \tau_{m}^{\mathrm{hyb}} > \max\Big\{\underset{m}{\mathrm{max}}\,\, \tau_{m}^{\mathrm{fso}}+\underset{n}{\mathrm{max}}\,\, \tau_{n}^{\mathrm{rf}}, \underset{(m,n)}{\mathrm{max}}\,\, \tau_{mn}^{\mathrm{mix}} \Big\} \\
\text{Case 2 (Independent Mode):} \\
 \quad\underset{m}{\mathrm{max}}\,\, \tau_{m}^{\mathrm{fso}}+\underset{n}{\mathrm{max}}\,\, \tau_{n}^{\mathrm{rf}} > \max\Big\{\underset{m}{\mathrm{max}}\,\, \tau_{m}^{\mathrm{hyb}}, \underset{(m,n)}{\mathrm{max}}\,\, \tau_{mn}^{\mathrm{mix}}\Big\} \\
\text{Case 3 (Mixed Mode):} \\
\quad\underset{(m,n)}{\mathrm{max}}\,\, \tau_{mn}^{\mathrm{mix}} > \max\Big\{\underset{m}{\mathrm{max}}\,\, \tau_{m}^{\mathrm{hyb}}, \underset{m}{\mathrm{max}}\,\, \tau_{m}^{\mathrm{fso}}+\underset{n}{\mathrm{max}}\,\, \tau_{n}^{\mathrm{rf}} \Big\}
\end{cases}
}\,\,\,
\end{IEEEeqnarray}

Using the RF and FSO relay selection and RF time allocation policies in (\ref{Eq:DL-theo}), the maximum throughput achieved by the protocol in Theorem~\ref{Theo:Delay-Limited}, denoted by $\tau^*$, is given by
\begin{IEEEeqnarray}{rll} \label{Eq:NBAOptThr}
\Scale[1]{\tau^* = \begin{cases}
\underset{m}{\mathrm{max}}\,\, \tau_{m}^{\mathrm{hyb}}&\mathrm{for}\,\,\text{Case 1} \\
\underset{m}{\mathrm{max}}\,\, \tau_{m}^{\mathrm{fso}}+\underset{n}{\mathrm{max}}\,\, \tau_{n}^{\mathrm{rf}}&\mathrm{for}\,\,\text{Case 2}\\
\underset{(m,n)}{\mathrm{max}}\,\, \tau_{mn}^{\mathrm{mix}}&\mathrm{for}\,\,\text{Case 3}
\end{cases}}
\end{IEEEeqnarray}
}{%
\begin{IEEEeqnarray}{lll}\label{Eq:DL-theo}
 \Scale[0.85]{ \alpha_{lm^*} \Equal }  \IEEEyesnumber\IEEEyessubnumber \\
  \Scale[0.85]{\begin{cases}
1,  &\mathrm{if}\,\, \big\{ \text{Case 1} \,\,\wedge\,\, m^* = \underset{m}{\mathrm{argmax}}\,\, \tau_{m}^{\mathrm{hyb}} \big\}\\
&\vee \,\, \big\{ \text{Case 2} \,\,\wedge\,\,  m^*  = \underset{m}{\mathrm{argmax}}\,\, \tau_{m}^{\mathrm{fso}} \big\}\\
&\vee \,\, \big\{ \text{Case 3} \, \wedge\,  l = 1 \, \wedge\, (m^*,-) = \underset{(m,n)}{\mathrm{argmax}}\,\, \tau_{mn}^{\mathrm{mix}} \big\}\\
&\vee \,\, \big\{ \text{Case 3} \, \wedge\,  l = 2 \, \wedge\, (-,m^*) = \underset{(m,n)}{\mathrm{argmax}}\,\, \tau_{mn}^{\mathrm{mix}} \big\}\\
0, &\mathrm{otherwise}
\end{cases} } \nonumber 
\\
 \Scale[0.85]{\beta_{lm^*} \Equal} \IEEEyessubnumber \\
 \Scale[0.85]{\begin{cases}
1,  &\mathrm{if}\,\, \big\{ \text{Case 1} \,\,\wedge\,\, m^* = \underset{m}{\mathrm{argmax}}\,\, \tau_{m}^{\mathrm{hyb}}  \big\}\\
&\vee \,\, \big\{ \text{Case 2} \,\,\wedge\,\, m^* = \underset{m}{\mathrm{argmax}}\,\, \tau_{m}^{\mathrm{rf}} \big\}\\
&\vee \,\, \big\{ \text{Case 3} \, \wedge\,  l = 1 \, \wedge\, (-,m^*) = \underset{(m,n)}{\mathrm{argmax}}\,\, \tau_{mn}^{\mathrm{mix}}  \big\}\\
&\vee \,\, \big\{ \text{Case 3} \, \wedge\,  l = 2 \, \wedge\, (m^*,-) = \underset{(m,n)}{\mathrm{argmax}}\,\, \tau_{mn}^{\mathrm{mix}}  \big\}\\
0, &\mathrm{otherwise}
\end{cases}}  \nonumber \\
\Scale[0.85]{ \rho_1^* \Equal 1 \Minus \rho_2^* \Equal}  \IEEEyessubnumber \\ 
 \Scale[0.85]{
  \begin{cases}
\left[ \frac{C_{2m^*}^{\mathrm{fso}} +C_{2m^*}^{\mathrm{rf}} -C_{1m^*}^{\mathrm{fso}} }{C_{1m^*}^{\mathrm{rf}} +C_{2m^*}^{\mathrm{rf}} } \right]_0^1, 
&  \mathrm{if}\,\, \big\{ \text{Case 1} \,\,\wedge\,\, m^* = \underset{m}{\mathrm{argmax}}\,\, \tau_{m}^{\mathrm{hyb}}  \big\}\\
 \frac{C_{2m^*}^{\mathrm{rf}} }{C_{1m^*}^{\mathrm{rf}} +C_{2m^*}^{\mathrm{rf}} }, \,\,  &\mathrm{if}\,\, \big\{ \text{Case 2} \,\,\wedge\,\, m^* = \underset{m}{\mathrm{argmax}}\,\, \tau_{m}^{\mathrm{rf}}  \big\}\\
\frac{C_{2m^*}^{\mathrm{fso}} }{C_{1m^*}^{\mathrm{rf}} }, \qquad &\mathrm{if}\,\, \big\{ \text{Case 3} \,\,\wedge\,\, (-,m^*) = \underset{(m,n)}{\mathrm{argmax}}\,\, \tau_{mn}^{\mathrm{mix}}  \big\}\\
 \end{cases} }   \nonumber 
\end{IEEEeqnarray}
where $\tau_{m}^{\mathrm{hyb}}$,  $\tau_{mn}^{\mathrm{ind}}$, and $\tau_{mn}^{\mathrm{mix}}$ are given by
\begin{IEEEeqnarray}{lll} \label{Eq:DiffThr}
\Scale[0.85]{ \tau_m^{\mathrm{hyb}}=\begin{cases}
C_{2m}^{\mathrm{fso}}+C_{2m}^{\mathrm{rf}}, \qquad \mathrm{if}\,\, \frac{C_{2m}^{\mathrm{fso}}+C_{2m}^{\mathrm{rf}}}{ C_{1m}^{\mathrm{fso}}}<1 \\
C_{1m}^{\mathrm{fso}}+C_{1m}^{\mathrm{rf}}, \qquad  \mathrm{if}\,\, \frac{C_{1m}^{\mathrm{fso}}+C_{1m}^{\mathrm{rf}}}{C_{2m}^{\mathrm{fso}}}<1 \\
\frac{C_{2m}^{\mathrm{fso}}+C_{2m}^{\mathrm{rf}}-C_{1m}^{\mathrm{fso}}}{C_{1m}^{\mathrm{rf}}+C_{2m}^{\mathrm{rf}}} C_{1m}^{\mathrm{rf}} + C_{1m}^{\mathrm{fso}},   \,\, \mathrm{otherwise}
\end{cases} }\IEEEyesnumber\IEEEyessubnumber \\
\Scale[0.85]{\tau_{mn}^{\mathrm{ind}}  =  \tau_{m}^{\mathrm{fso}}  + \tau_{n}^{\mathrm{rf}},\,\,\,  \mathrm{where} \,\,\,
\begin{cases} \tau_{m}^{\mathrm{fso}} = \min\big\{C_{1m}^{\mathrm{fso}},C_{2m}^{\mathrm{fso}}\big\} \\  \tau_{n}^{\mathrm{rf}} = \frac{C_{1n}^{\mathrm{rf}} C_{2n}^{\mathrm{rf}}}{C_{1n}^{\mathrm{rf}}+C_{2n}^{\mathrm{rf}}} \end{cases}}   \IEEEyessubnumber \\
\Scale[0.85]{\tau_{mn}^{\mathrm{mix}} =  \begin{cases}
C_{1m}^{\mathrm{fso}}+C_{2n}^{\mathrm{fso}},\,\,\,\,&\mathrm{if}\,\,\frac{C_{2n}^{\mathrm{fso}}}{C_{1n}^{\mathrm{rf}}}+\frac{C_{1m}^{\mathrm{fso}}}{C_{2m}^{\mathrm{rf}}}\leq 1 \\
0,&\mathrm{otherwise.}
\end{cases}}   \IEEEyessubnumber  
\end{IEEEeqnarray}
 Moreover, Cases 1-3 are defined as follows
\begin{IEEEeqnarray}{rll} \label{Eq:Cases}
\Scale[0.85]{
\begin{cases}
\text{Case 1 (Hybrid Mode):} \\
\quad \underset{m}{\mathrm{max}}\,\, \tau_{m}^{\mathrm{hyb}} > \max\Big\{\underset{m}{\mathrm{max}}\,\, \tau_{m}^{\mathrm{fso}}+\underset{n}{\mathrm{max}}\,\, \tau_{n}^{\mathrm{rf}}, \underset{(m,n)}{\mathrm{max}}\,\, \tau_{mn}^{\mathrm{mix}} \Big\} \\
\text{Case 2 (Independent Mode):} \\
 \quad\underset{m}{\mathrm{max}}\,\, \tau_{m}^{\mathrm{fso}}+\underset{n}{\mathrm{max}}\,\, \tau_{n}^{\mathrm{rf}} > \max\Big\{\underset{m}{\mathrm{max}}\,\, \tau_{m}^{\mathrm{hyb}}, \underset{(m,n)}{\mathrm{max}}\,\, \tau_{mn}^{\mathrm{mix}}\Big\} \\
\text{Case 3 (Mixed Mode):} \\
\quad\underset{(m,n)}{\mathrm{max}}\,\, \tau_{mn}^{\mathrm{mix}} > \max\Big\{\underset{m}{\mathrm{max}}\,\, \tau_{m}^{\mathrm{hyb}}, \underset{m}{\mathrm{max}}\,\, \tau_{m}^{\mathrm{fso}}+\underset{n}{\mathrm{max}}\,\, \tau_{n}^{\mathrm{rf}} \Big\}
\end{cases}
}\,\,\,
\end{IEEEeqnarray}

Using the RF and FSO relay selection and RF time allocation policies in (\ref{Eq:DL-theo}), the maximum throughput achieved by the protocol in Theorem~\ref{Theo:Delay-Limited}, denoted by $\tau^*$, is given by
\begin{IEEEeqnarray}{rll} \label{Eq:NBAOptThr}
\Scale[0.85]{\tau^* = \begin{cases}
\underset{m}{\mathrm{max}}\,\, \tau_{m}^{\mathrm{hyb}}&\mathrm{for}\,\,\text{Case 1} \\
\underset{m}{\mathrm{max}}\,\, \tau_{m}^{\mathrm{fso}}+\underset{n}{\mathrm{max}}\,\, \tau_{n}^{\mathrm{rf}}&\mathrm{for}\,\,\text{Case 2}\\
\underset{(m,n)}{\mathrm{max}}\,\, \tau_{mn}^{\mathrm{mix}}&\mathrm{for}\,\,\text{Case 3}
\end{cases}}
\end{IEEEeqnarray}
}
 \end{theo}
\begin{IEEEproof}
Please refer to Appendix \ref{App:Theo2}.  
\end{IEEEproof}

The feasible  sets $\boldsymbol{\mathcal{A}}$ and $\boldsymbol{\mathcal{B}}$ of the relay selection variables allow the selection of at most four different relays for RF/FSO reception and transmission. However, due to the constraints in (\ref{Eq:Opt_DL}), the optimal relay selection policy selects at most two different relays for RF/FSO reception and transmission in order to ensure that the data which is transmitted from the source to a certain relay can be actually forwarded to the destination. Moreover, the optimal throughput maximizing policy in Theorem~\ref{Theo:Delay-Limited} reveals that the optimal relay selection policy $(\alpha_{lm},\beta_{lm})$ belongs to one of the following three cases, see Fig.~\ref{Fig:Mode_DL}.
\\ \textit{Case 1 (Hybrid Mode):}  The same relay $\mathcal{R}_m$ is selected for  FSO/RF transmission/reception, i.e., the RF
links serve as support links for the FSO links.\\
\textit{Case 2 (Independent Mode):} Relay $\mathcal{R}_m$ is selected for FSO reception and transmission and a different relay $\mathcal{R}_n$ is selected for  RF reception and transmission, i.e., the FSO and RF links are used independently.\\
\textit{Case 3 (Mixed Mode):} Relays $\mathcal{R}_m$ and $\mathcal{R}_n$, $m \neq n$, are selected for FSO reception and transmission, respectively, and relays $\mathcal{R}_m$ and $\mathcal{R}_n$ are selected for RF transmission and reception, respectively.

\iftoggle{OneColumn}{%
\begin{figure}
\centering
\scalebox{0.8}{
\pstool[width=1\linewidth]{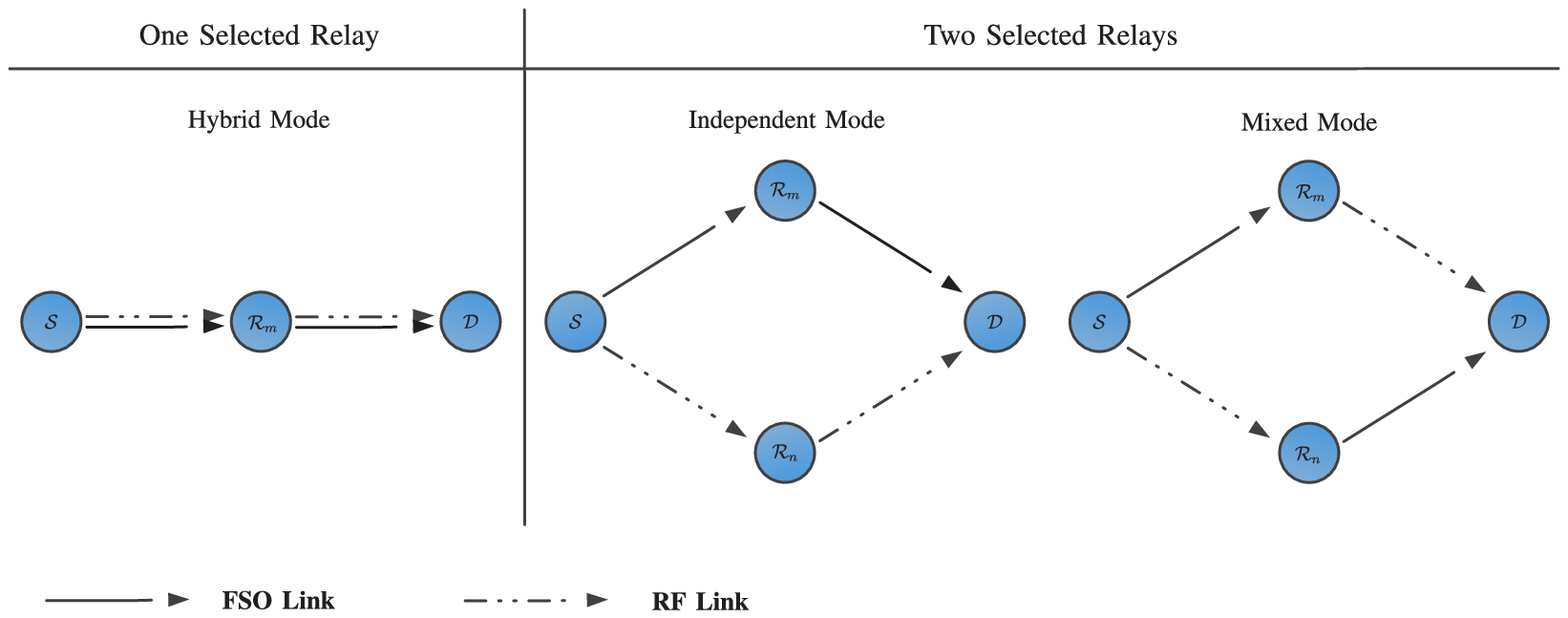}{
\psfrag{A}[c][c][0.5]{$\mathcal{S}$}
\psfrag{B}[c][c][0.5]{$\mathcal{D}$}
\psfrag{m}[c][c][0.5]{$\mathcal{R}_m$}
\psfrag{n}[c][c][0.5]{$\mathcal{R}_n$}
\psfrag{RF}[c][r][0.65]{$\textbf{RF Link}$}
\psfrag{FS}[c][r][0.65]{$\textbf{FSO Link}$}
\psfrag{C1}[c][c][0.65]{$\text{Hybrid Mode}$}
\psfrag{C2}[c][c][0.65]{$\text{Independent Mode}$}
\psfrag{C3}[c][c][0.65]{$\text{Mixed Mode}$}
\psfrag{R1}[c][c][0.8]{$\text{One Selected Relay}$}
\psfrag{R2}[c][c][0.8]{$\text{Two Selected Relays}$}
}}
\vspace{+0.3cm}
\caption{The three  possible optimal non-BA relaying modes in the considered parallel hybrid RF/FSO relay channel.}
\label{Fig:Mode_DL}\vspace{-0.3cm}
\end{figure}
}{%
\begin{figure*}
\centering
\scalebox{0.75}{
\pstool[width=1\linewidth]{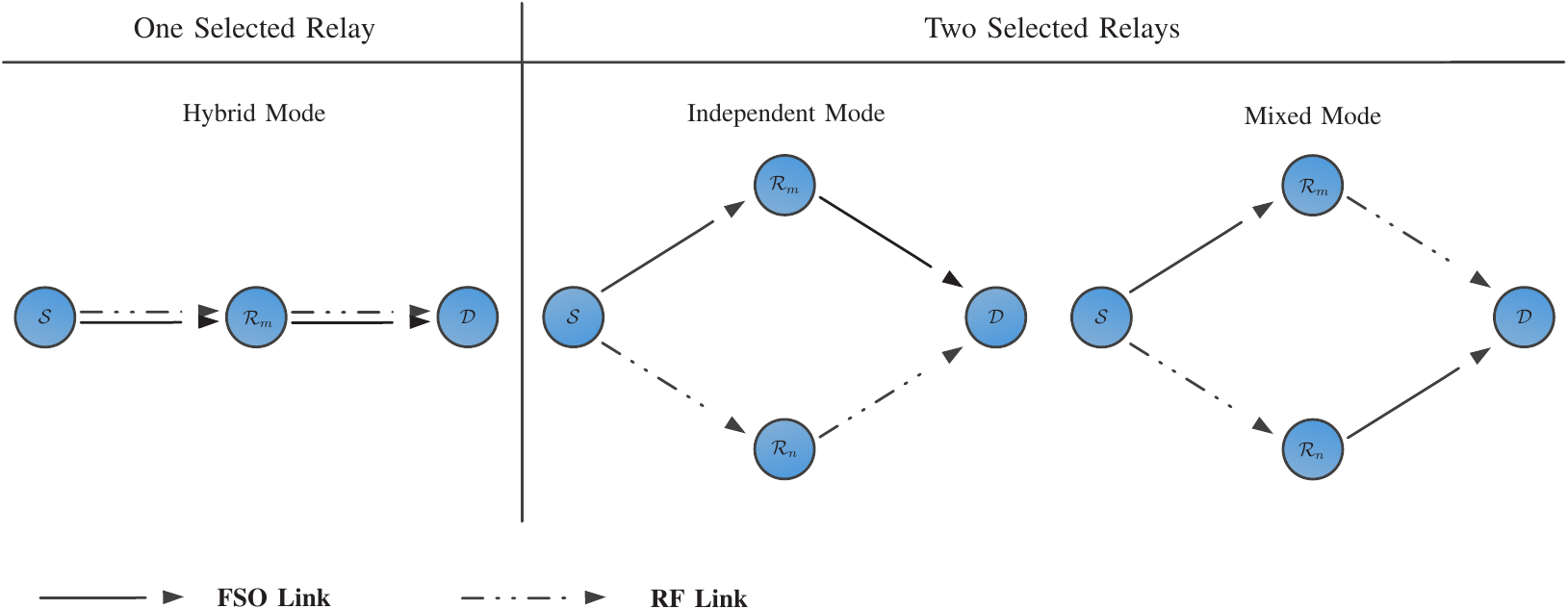}{
\psfrag{A}[c][c][0.5]{$\mathcal{S}$}
\psfrag{B}[c][c][0.5]{$\mathcal{D}$}
\psfrag{m}[c][c][0.5]{$\mathcal{R}_m$}
\psfrag{n}[c][c][0.5]{$\mathcal{R}_n$}
\psfrag{RF}[c][r][0.65]{$\textbf{RF Link}$}
\psfrag{FS}[c][r][0.65]{$\textbf{FSO Link}$}
\psfrag{C1}[c][c][0.65]{$\text{Hybrid Mode}$}
\psfrag{C2}[c][c][0.65]{$\text{Independent Mode}$}
\psfrag{C3}[c][c][0.65]{$\text{Mixed Mode}$}
\psfrag{R1}[c][c][0.8]{$\text{One Selected Relay}$}
\psfrag{R2}[c][c][0.8]{$\text{Two Selected Relays}$}
}}
\vspace{+0.3cm}
\caption{The three  possible optimal non-BA relaying modes in the considered parallel hybrid RF/FSO relay channel.}
\label{Fig:Mode_DL}\vspace{-0.3cm}
\end{figure*}
}

The optimal transmission time allocation to the RF links given in~(\ref{Eq:DL-theo}c) is found such that the bottleneck throughput of  the $\mathcal{S}-\mathcal{R}_m$ and $\mathcal{R}_m-\mathcal{D}$ links is maximized. Thereby, depending on whether $\mathcal{R}_m$ uses both the RF and FSO links, i.e., the hybrid and mixed modes, or only the RF links, i.e., the independent mode, the resulting optimal RF time allocation  policy depends on both the RF and FSO fading states or only the RF fading state, respectively. 

\subsection{Optimal BA Policy}
 In this subsection, we assume that the relay nodes take  advantage of buffering to transmit/receive  in each  time slot over the RF/FSO links which have the best qualities. We assume that each relay is equipped with an infinite-size buffer for data storage. Let $Q_m[b],\,\,m\in\{1,\dots,M\}$, denote the amount of information in bits available in the buffer of relay $m$ at the end of the $b$-th time slot. The dynamics of the queues at the relay nodes can be modelled as
 \iftoggle{OneColumn}{%
\begin{IEEEeqnarray}{cll}\label{Eq:Q_Unlimit}
Q_{m}[b]=Q_{m}[b-1]&+ \underset{R_{1m}^{\mathrm{fso}}[b]}{\underbrace{\alpha_{1m}[b]  C_{1m}^{\mathrm{fso}}[b]}} +\underset{R_{1m}^{\mathrm{rf}}[b]}{\underbrace{\beta_{1m}[b] \rho_1[b]  C_{1m}^{\mathrm{rf}}[b]}} 
-\underset{R_{2m}^{\mathrm{fso}}[b]}{\underbrace{\min\big\{Q_m[b-1],\alpha_{2m}[b]  C_{2m}^{\mathrm{fso}}[b]\big\}}}  \\
&-\underset{R_{2m}^{\mathrm{rf}}[b]}{\underbrace{\min\Big\{\big[Q_m[b-1]-\alpha_{2m}[b]  C_{2m}^{\mathrm{fso}}[b]\big]^+ 
+\beta_{1m}[b] \rho_1[b]  C_{1m}^{\mathrm{rf}}[b], \beta_{2m}[b] \rho_2[b]  C_{2m}^{\mathrm{rf}}[b]\Big\}}},\nonumber
\end{IEEEeqnarray}
}{%
\begin{IEEEeqnarray}{lll}\label{Eq:Q_Unlimit}
Q_{m}[b]=&Q_{m}[b-1] \\
&+ \underset{R_{1m}^{\mathrm{fso}}[b]}{\underbrace{\alpha_{1m}[b]  C_{1m}^{\mathrm{fso}}[b]}} +\underset{R_{1m}^{\mathrm{rf}}[b]}{\underbrace{\beta_{1m}[b] \rho_1[b]  C_{1m}^{\mathrm{rf}}[b]}} \nonumber \\
&-\underset{R_{2m}^{\mathrm{fso}}[b]}{\underbrace{\min\big\{Q_m[b-1],\alpha_{2m}[b]  C_{2m}^{\mathrm{fso}}[b]\big\}}} \nonumber \\
&-\underbrace{\min\Big\{\big[Q_m[b-1]-\alpha_{2m}[b]  C_{2m}^{\mathrm{fso}}[b]\big]^+} \nonumber \\
&\overset{R_{2m}^{\mathrm{rf}}[b]}{\overbrace{+\beta_{1m}[b] \rho_1[b]  C_{1m}^{\mathrm{rf}}[b], \beta_{2m}[b] \rho_2[b]  C_{2m}^{\mathrm{rf}}[b]\Big\}}},\nonumber
\end{IEEEeqnarray}
}
where $R_{1m}^{\mathrm{fso}}[b]$, $R_{1m}^{\mathrm{rf}}[b]$, $R_{2m}^{\mathrm{fso}}[b]$, and $R_{2m}^{\mathrm{rf}}[b]$ are the data rates of the $\mathcal{S}-\mathcal{R}_m$ FSO, $\mathcal{S}-\mathcal{R}_m$ RF, $\mathcal{R}_m-\mathcal{D}$ FSO, and $\mathcal{R}_m-\mathcal{D}$ RF links, respectively, in the $b$-th time slot.
In particular, at the beginning of each time slot, the amount of data sent over the $\mathcal{R}_m-\mathcal{D}$ FSO link is limited by the capacity of the $\mathcal{R}_m-\mathcal{D}$ FSO link, i.e., $\alpha_{2m}[b]  C_{2m}^{\mathrm{fso}}[b]$, and the amount of information available at the relay's buffer, i.e., $Q_m[b-1]$. Similarly, in the second half of the time slot, the amount of data used by relay $\mathcal{R}_m$ to encode the RF codewords is limited by the capacity of the $\mathcal{R}_m-\mathcal{D}$ RF link, i.e, $\beta_{2m}[b] \rho_2[b]  C_{2m}^{\mathrm{rf}}[b]$, and the the amount of information in the buffer, i.e., $\big[Q_m[b-1]-\alpha_{2m}[b]  C_{2m}^{\mathrm{fso}}[b]\big]^+ 
+\beta_{1m}[b] \rho_1[b]  C_{1m}^{\mathrm{rf}}[b]$.
Since the throughput is equal to the amount of data that is received at the destination, the throughput maximization problem for the BA relaying protocol can be written as
\begin{IEEEeqnarray}{cll}\label{Eq:Opt_DUL_Queue}
\underset{\boldsymbol{\alpha}\in{{\boldsymbol{\mathcal{A}}}},\boldsymbol{\beta}\in{{\boldsymbol{\mathcal{B}}}},\boldsymbol{\rho}\in{{\boldsymbol{\mathcal{C}}}}}{\mathrm{maximize}}\,\,&\bar{\tau}=\sum_{\forall m}\bar{\tau}_m = \sum_{\forall m} \big[ R_{2m}^{\mathrm{fso}}[b]+R_{2m}^{\mathrm{rf}}[b] \big]. \quad
\end{IEEEeqnarray}
Solving the above optimization problem is quite involved due to recursive dynamics of the queue (\ref{Eq:Q_Unlimit}) which appear in $R_{2m}^{\mathrm{fso}}[b]$ and $R_{2m}^{\mathrm{rf}}[b]$. To tackle this problem, we use a useful result from  queuing theory \cite[Chapter~2]{Neely}, \cite[Eq. (50)]{ITIEEE}. Suppose $\mathtt{A}[b]$, $\mathtt{D}[b]$, $\mathtt{C}[b]$, and $\mathtt{Q}[b]$ are the arrival rate, the departure rate, the processing rate (departure capacity), and the amount of information  of a queue in the $b$-th time slot, respectively. Thereby, although the \textit{instantaneous} departure rate of the queue is limited by the amount of data available at the queue, i.e., $\mathtt{D}[b]=\min\{\mathtt{Q}[b],\mathtt{C}[b]\}$, the \textit{average} departure rate can be written independent of the dynamics of the queue as $\mathbbmss{E}\{\mathtt{D}\}=\min\big\{\mathbbmss{E}\{\mathtt{A}\},\mathbbmss{E}\{\mathtt{C}\}\big\}$, see \cite[Appendix~E]{ITIEEE} for a detailed proof. Using this result and as $B\to\infty$, the throughput maximizing policy for this case can be obtained by solving the following optimization problem
\iftoggle{OneColumn}{%
\begin{IEEEeqnarray}{cll}\label{Eq:Opt_DUL}
\underset{\boldsymbol{\alpha}\in{{\boldsymbol{\mathcal{A}}}},\boldsymbol{\beta}\in{{\boldsymbol{\mathcal{B}}}},\boldsymbol{\rho}\in{{\boldsymbol{\mathcal{C}}}},\boldsymbol{\tau}\geq \mathbf{0}}{\mathrm{maximize}}\,\,&\bar{\tau}=\sum_{\forall m}\bar{\tau}_m  \\
\mathrm{subject~to} \,\, & \bar{\tau}_m \leq\dfrac{1}{B} \sum_{\forall b} \left[ \alpha_{1m}[b]  C_{1m}^{\mathrm{fso}}[b] +\beta_{1m}[b] \rho_1[b]  C_{1m}^{\mathrm{rf}}[b]\right], \quad \forall m, \nonumber \\
&\bar{\tau}_m \leq \dfrac{1}{B} \sum_{\forall b} \left[ \alpha_{2m}[b]  C_{2m}^{\mathrm{fso}}[b] +\beta_{2m}[b] \rho_2[b]  C_{2m}^{\mathrm{rf}}[b]\right], \quad \forall m,\nonumber
\end{IEEEeqnarray}
}{%
\begin{IEEEeqnarray}{cll}\label{Eq:Opt_DUL}
\underset{\boldsymbol{\alpha}\in{{\boldsymbol{\mathcal{A}}}},\boldsymbol{\beta}\in{{\boldsymbol{\mathcal{B}}}},\boldsymbol{\rho}\in{{\boldsymbol{\mathcal{C}}}},\boldsymbol{\tau}\geq \mathbf{0}}{\mathrm{maximize}}\,\,&\bar{\tau}=\sum_{\forall m}\bar{\tau}_m  \\
\mathrm{subject~to} \,\, & \bar{\tau}_m \leq\dfrac{1}{B} \sum_{\forall b} \big[ \alpha_{1m}[b]  C_{1m}^{\mathrm{fso}}[b] \nonumber \\
&\qquad\qquad+\beta_{1m}[b] \rho_1[b]  C_{1m}^{\mathrm{rf}}[b]\big], \quad \forall m, \nonumber \\
&\bar{\tau}_m \leq \dfrac{1}{B} \sum_{\forall b} \big[ \alpha_{2m}[b]  C_{2m}^{\mathrm{fso}}[b] \nonumber \\
&\qquad\qquad +\beta_{2m}[b] \rho_2[b]  C_{2m}^{\mathrm{rf}}[b]\big], \quad \forall m,\nonumber
\end{IEEEeqnarray}
}
where the right-hand sides of the first and second constraints are the average arrival rate and the average departure capacity of the queue at $\mathcal{R}_m$, respectively. 

As can be observed from the constraints  in (\ref{Eq:Opt_DUL}),  for  BA relaying, the \textit{average} throughput of each relay is limited.  In contrast, for non-BA relaying, cf.  (\ref{Eq:Opt_DL}), the \textit{instantaneous} throughput of each relay is limited. Therefore, the feasible set of the problem in (\ref{Eq:Opt_DUL}) is larger than that of (\ref{Eq:Opt_DL}) which leads to a higher achievable throughput for the BA relaying protocol. The higher achievable throughput of the BA protocol comes at the expense of  an increased end-to-end delay. Hence, the BA protocol is a suitable option for delay-tolerant applications. In the following theorem, we present the optimal BA relay selection policy as the solution of the problem in (\ref{Eq:Opt_DUL}). For notational simplicity, let $C_{lm}^{\mathrm{fso}}(h_{lm})$ and $C_{lm}^{\mathrm{rf}}(g_{lm})$ denote the capacities of the FSO and RF links as  functions of the  fading states, respectively. Moreover, $f_{\boldsymbol{h}_l}(\boldsymbol{h}_l)$ and $f_{\boldsymbol{g}_l}(\boldsymbol{g}_l), \l=1,2,$ denote the  pdfs of the  
random vectors $\boldsymbol{h}_l$ and $\boldsymbol{g}_l$, respectively, where $\boldsymbol{h}_l$ and $\boldsymbol{g}_l$ are the vectors containing the fading coefficients of the $l$-th hop of the FSO links and the RF links, respectively.
 Furthermore, we introduce constant vector $\boldsymbol{\lambda}=[\lambda_1, \lambda_2,\dots,\lambda_M]$ which we will use for the statement of the optimal protocol. The elements of vector $\boldsymbol{\lambda}$  are in fact related to
 the Lagrange multipliers corresponding to the constraints in (\ref{Eq:Opt_DUL}).

\begin{theo}\label{Theo:Delay-Unlimited}
For the  parallel  BA relay channel with hybrid RF/FSO links, the optimal policies for FSO and RF relay selection and optimal RF transmission time allocation as a function of the fading state are given by
\begin{IEEEeqnarray}{rll}\label{Eq:OptSelec}
\alpha_{lm^*}(\boldsymbol{h}_l)&={\begin{cases}
1, \quad &\mathrm{if}\,\, m^* = \underset{m}{\mathrm{argmax}}\,\, \Lambda_{lm}^{\mathrm{fso}}({h}_{lm})\\
0,\quad &\mathrm{otherwise},
\end{cases}} \quad \IEEEyesnumber\IEEEyessubnumber
\\
\beta_{lm^*}(\boldsymbol{g}_1,\boldsymbol{g}_2)&={\begin{cases}
1, \quad &\mathrm{if}\,\, m^* = \underset{l,m}{\mathrm{argmax}}\,\, \Lambda_{lm}^{\mathrm{rf}}({g}_{lm}) \\
0,\quad &\mathrm{otherwise},
\end{cases}} \IEEEyessubnumber\\
\rho_{l^*}(\boldsymbol{g}_1,\boldsymbol{g}_2)&={\begin{cases}
1, \quad &\mathrm{if}\,\, l^* = \underset{l,m}{\mathrm{argmax}}\,\, \Lambda_{lm}^{\mathrm{rf}}({g}_{lm}) \\
0,\quad &\mathrm{otherwise},
\end{cases}} \IEEEyessubnumber
\end{IEEEeqnarray}
where $\Lambda_{1m}^{\mathrm{fso}}({h}_{1m})=\lambda_m C_{1m}^{\mathrm{fso}}({h}_{1m})$, $\Lambda_{2m}^{\mathrm{fso}}({h}_{2m})=(1-\lambda_m)C_{2m}^{\mathrm{fso}}({h}_{2m})$, $\Lambda_{1m}^{\mathrm{rf}}({g}_{1m})=\lambda_m C_{1m}^{\mathrm{rf}}({g}_{1m})$, and $\Lambda_{2m}^{\mathrm{rf}}({g}_{2m})=(1-\lambda_m) C_{2m}^{\mathrm{rf}}({g}_{2m})$. In addition, $\lambda_m$ is a constant which depends on the fading distributions $f_{\boldsymbol{h}_l}(\boldsymbol{h}_l)$ and $f_{\boldsymbol{g}_l}(\boldsymbol{g}_l)$. The optimal value of $\lambda_m$ can be obtained offline before transmission starts using an iterative algorithm with the following update equation in the $k$-th iteration
\iftoggle{OneColumn}{%
\begin{IEEEeqnarray}{cll}\label{Eq:Upd}
{\lambda_m}[k+1]=\Big[  {\lambda_m}[k]-\epsilon_m[k]\left(\bar{C}_{1m}^{\mathrm{fso}}[k] +\bar{C}_{1m}^{\mathrm{rf}}[k] -\bar{C}_{2m}^{\mathrm{fso}}[k] -\bar{C}_{2m}^{\mathrm{rf}}[k] \right)  \Big]_0^1,
 \end{IEEEeqnarray}
}{%
\begin{IEEEeqnarray}{cll}\label{Eq:Upd}
{\lambda_m}[k+1]=\Big[  {\lambda_m}[k]-\epsilon_m[k]\big(&\bar{C}_{1m}^{\mathrm{fso}}[k] +\bar{C}_{1m}^{\mathrm{rf}}[k] \nonumber \\
&-\bar{C}_{2m}^{\mathrm{fso}}[k] -\bar{C}_{2m}^{\mathrm{rf}}[k] \big)  \Big]_0^1, \qquad
 \end{IEEEeqnarray}
}
 where  $\epsilon_m[k],\,\forall m,$ is a sufficiently small step size. Moreover, the average capacity terms, $\bar{C}_{lm}^{\mathrm{fso}}[k]$ and $\bar{C}_{lm}^{\mathrm{rf}}[k]$, are given by
 \iftoggle{OneColumn}{%
 \begin{IEEEeqnarray}{rll}\label{Eq:C_Avg}
  \bar{C}_{lm}^{\mathrm{fso}}[k]&=\mathbbmss{E}\left\{\alpha_{lm^*}(\boldsymbol{h}_l)C_{lm}^{\mathrm{fso}}({h}_{lm})\right\}
  =\int_{\boldsymbol{h}_l}\alpha_{lm^*}(\boldsymbol{h}_l)C_{lm}^{\mathrm{fso}}({h}_{lm})f_{\boldsymbol{h}_l}(\boldsymbol{h}_l) \mathrm{d}\boldsymbol{h}_l,\quad l=1,2,
  \IEEEyesnumber\IEEEyessubnumber
\\
\bar{C}_{lm}^{\mathrm{rf}}[k]&=\mathbbmss{E}\left\{\beta_{lm^*}(\boldsymbol{g}_1,\boldsymbol{g}_2) \rho_{l^*}(\boldsymbol{g}_1,\boldsymbol{g}_2) C_{lm}^{\mathrm{rf}}({g}_{lm})\right\}
  \nonumber\\
  &=\iint_{\boldsymbol{g}_1,\boldsymbol{g}_2} \beta_{lm^*}(\boldsymbol{g}_1,\boldsymbol{g}_2) \rho_{l^*}(\boldsymbol{g}_1,\boldsymbol{g}_2) C_{lm}^{\mathrm{rf}}({g}_{lm}) f_{\boldsymbol{g}_1}(\boldsymbol{g}_1) f_{\boldsymbol{g}_2}(\boldsymbol{g}_2)  \mathrm{d}\boldsymbol{g}_1 \mathrm{d}\boldsymbol{g}_2,\quad l=1,2,\IEEEyessubnumber
\end{IEEEeqnarray}
}{%
 \begin{IEEEeqnarray}{lll}\label{Eq:C_Avg}
  \bar{C}_{lm}^{\mathrm{fso}}[k]&=\mathbbmss{E}\left\{\alpha_{lm^*}(\boldsymbol{h}_l)C_{lm}^{\mathrm{fso}}({h}_{lm})\right\} \IEEEyesnumber\IEEEyessubnumber \\
 & =\int_{\boldsymbol{h}_l}\alpha_{lm^*}(\boldsymbol{h}_l)C_{lm}^{\mathrm{fso}}({h}_{lm})f_{\boldsymbol{h}_l}(\boldsymbol{h}_l) \mathrm{d}\boldsymbol{h}_l,\quad l=1,2,
  \nonumber
\\
\bar{C}_{lm}^{\mathrm{rf}}[k]&=\mathbbmss{E}\left\{\beta_{lm^*}(\boldsymbol{g}_1,\boldsymbol{g}_2) \rho_{l^*}(\boldsymbol{g}_1,\boldsymbol{g}_2) C_{lm}^{\mathrm{rf}}({g}_{lm})\right\}\IEEEyessubnumber\\
  &=\iint_{\boldsymbol{g}_1,\boldsymbol{g}_2} \beta_{lm^*}(\boldsymbol{g}_1,\boldsymbol{g}_2) \rho_{l^*}(\boldsymbol{g}_1,\boldsymbol{g}_2) C_{lm}^{\mathrm{rf}}({g}_{lm}) \nonumber \\
  & \qquad \qquad f_{\boldsymbol{g}_1}(\boldsymbol{g}_1) f_{\boldsymbol{g}_2}(\boldsymbol{g}_2)  \mathrm{d}\boldsymbol{g}_1 \mathrm{d}\boldsymbol{g}_2,\quad l=1,2,\nonumber
\end{IEEEeqnarray}
}
 where $\alpha_{lm^*}(\boldsymbol{h}_l)$, $\beta_{lm^*}(\boldsymbol{g}_1,\boldsymbol{g}_2)$, and $ \rho_{l^*}(\boldsymbol{g}_1,\boldsymbol{g}_2)$ are given by (\ref{Eq:OptSelec}) with $\lambda_m=\lambda_m[k]$.
 Substituting the optimal FSO and RF relay selection and RF time allocation variables from (\ref{Eq:OptSelec}) and  the optimal $\boldsymbol{\lambda}^*$ from (\ref{Eq:Upd}) into (\ref{Eq:C_Avg}), the maximum throughput is obtained as 
 \begin{IEEEeqnarray}{rll}\label{Eq:Thr_Opt_BA}
\bar{\tau}^*=\sum_m \bar{\tau}_m^*=\sum_m \mathrm{min}\left\{\bar{C}_{1m}^{\mathrm{fso}}+ \bar{C}_{1m}^{\mathrm{rf}}, \bar{C}_{2m}^{\mathrm{fso}} +\bar{C}_{2m}^{\mathrm{rf}}\right\}\hspace{-0.1cm}.\,\,\quad 
\end{IEEEeqnarray} 
\end{theo}
\begin{IEEEproof}  
Please refer to Appendix~B.
\end{IEEEproof} 

Recall that the optimal  non-BA protocol in Theorem~\ref{Theo:Delay-Limited} selects at most two different relays for RF/FSO reception and transmission. On the other hand,  exploiting the buffering capability of the relay nodes and   the degrees of freedom   available in the feasible  sets $\boldsymbol{\mathcal{A}}$ and $\boldsymbol{\mathcal{B}}$,  the BA protocol may select up to four different relays for RF/FSO reception and transmission in one time slot because the relays  are not forced  to  immediately forward the information received  from the source to the destination. However,  Theorem~\ref{Theo:Delay-Unlimited} reveals that it is optimal to select  at most three different relays, namely two relays for FSO reception and transmission and one relay for either  RF reception or transmission. The selection of only one relay for the RF links  leads to binary values for the  RF time allocation variable in (\ref{Eq:OptSelec}c), i.e., RF time allocation reduces to RF link selection.
Moreover, based on the number of  relays selected by the optimal BA protocol, there are ten possible transmission modes which  are illustrated in Fig.~\ref{Fig:Mode_DU}. These ten transmission modes can be further categorized into the following three types.
 \textit{i) Hybrid Modes:} The RF link is used as  backup for one of the  FSO links.
\textit{ii) Independent Modes:} None of the relays uses both RF and FSO links.
\textit{iii) Mixed Modes:} The RF link is cascaded with one of the FSO links.

\iftoggle{OneColumn}{%
\begin{figure}
\centering
\scalebox{0.8}{
\pstool[width=1\linewidth]{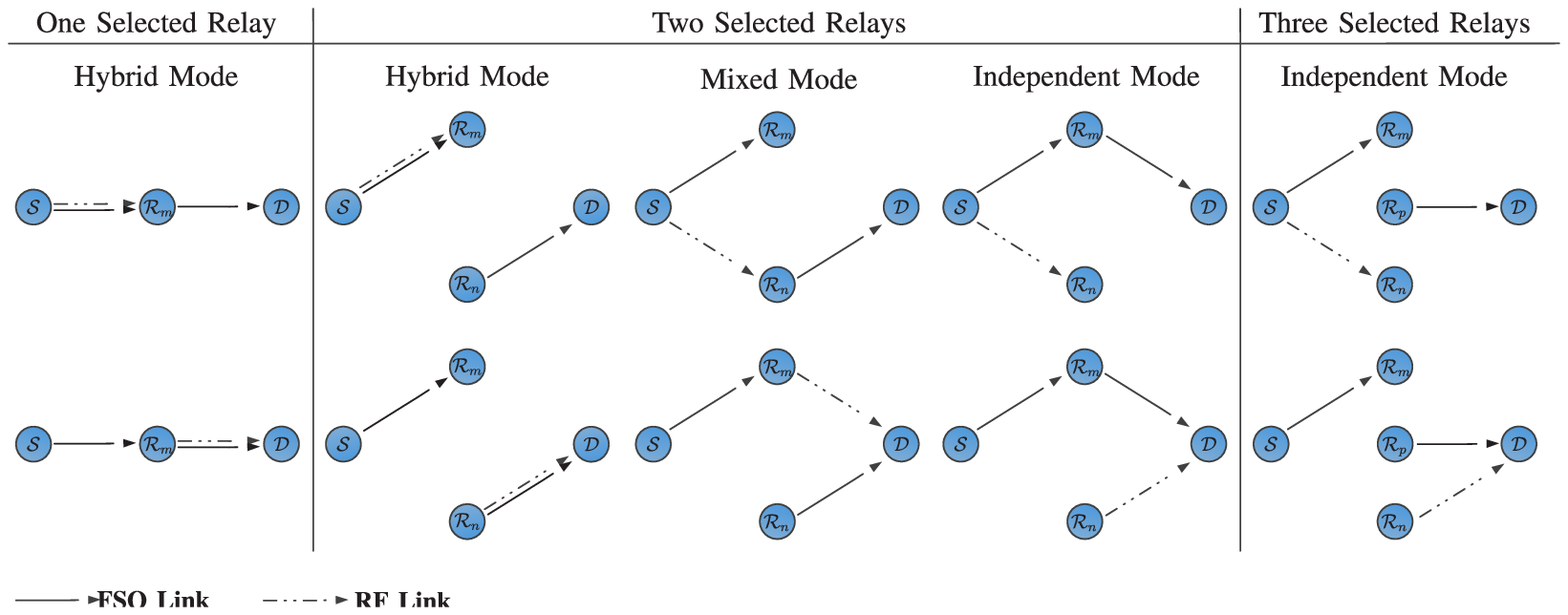}{
\psfrag{A}[c][c][0.5]{$\mathcal{S}$}
\psfrag{B}[c][c][0.5]{$\mathcal{D}$}
\psfrag{m}[c][c][0.5]{$\mathcal{R}_m$}
\psfrag{n}[c][c][0.5]{$\mathcal{R}_n$}
\psfrag{p}[c][c][0.5]{$\mathcal{R}_p$}
\psfrag{RF}[c][r][0.65]{$\textbf{RF Link}$}
\psfrag{FS}[c][r][0.65]{$\textbf{FSO Link}$}
\psfrag{C3}[c][c][0.75]{$\text{Independent Mode}$}
\psfrag{C1}[c][c][0.75]{$\text{Hybrid Mode}$}
\psfrag{C2}[c][c][0.75]{$\text{Mixed Mode}$}
\psfrag{R1}[c][c][0.8]{$\text{One Selected Relay}$}
\psfrag{R2}[c][c][0.8]{$\text{Two Selected Relays}$}
\psfrag{R3}[c][c][0.8]{$\text{Three Selected Relays}$}
}}
\vspace{+0.3cm}
\caption{The ten  possible optimal BA relaying modes in the considered parallel hybrid RF/FSO relay channel.}
\label{Fig:Mode_DU}\vspace{-0.3cm}
\end{figure}
}{%
\begin{figure*}
\centering
\scalebox{0.8}{
\pstool[width=1\linewidth]{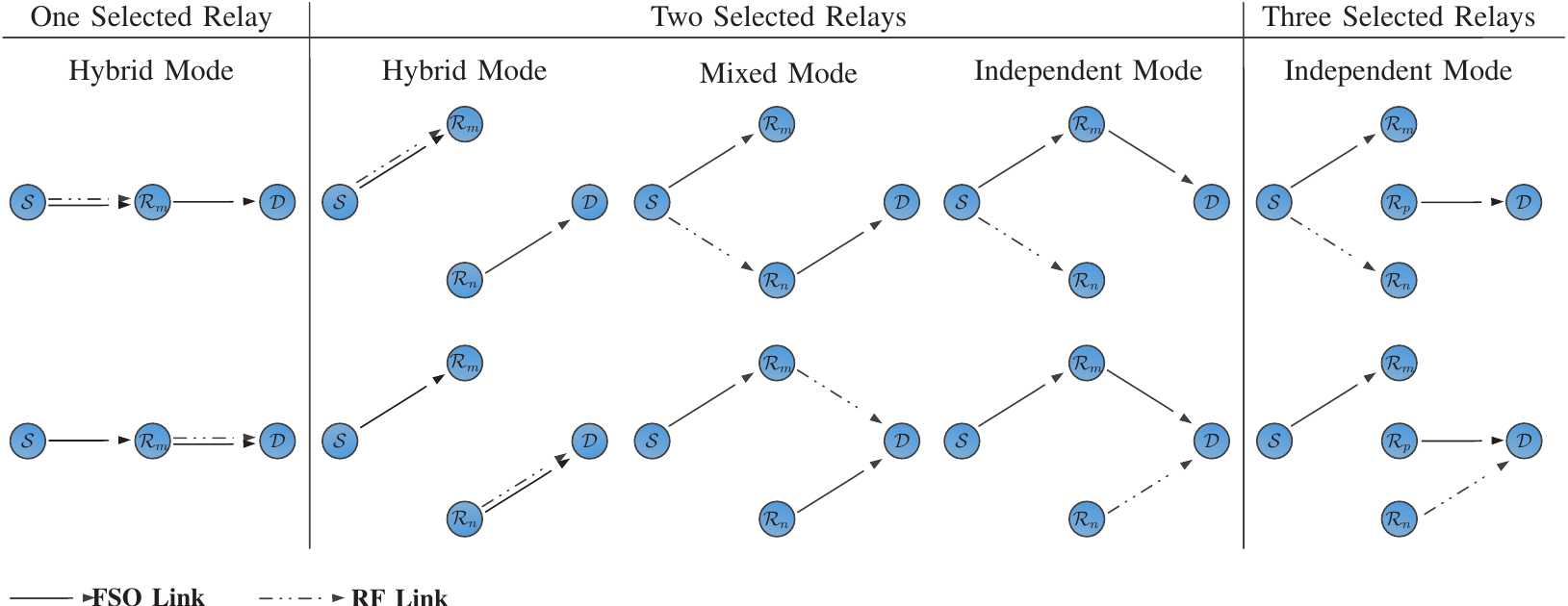}{
\psfrag{A}[c][c][0.5]{$\mathcal{S}$}
\psfrag{B}[c][c][0.5]{$\mathcal{D}$}
\psfrag{m}[c][c][0.5]{$\mathcal{R}_m$}
\psfrag{n}[c][c][0.5]{$\mathcal{R}_n$}
\psfrag{p}[c][c][0.5]{$\mathcal{R}_p$}
\psfrag{RF}[c][r][0.65]{$\textbf{RF Link}$}
\psfrag{FS}[c][r][0.65]{$\textbf{FSO Link}$}
\psfrag{C3}[c][c][0.75]{$\text{Independent Mode}$}
\psfrag{C1}[c][c][0.75]{$\text{Hybrid Mode}$}
\psfrag{C2}[c][c][0.75]{$\text{Mixed Mode}$}
\psfrag{R1}[c][c][0.8]{$\text{One Selected Relay}$}
\psfrag{R2}[c][c][0.8]{$\text{Two Selected Relays}$}
\psfrag{R3}[c][c][0.8]{$\text{Three Selected Relays}$}
}}
\vspace{+0.3cm}
\caption{The ten  possible optimal BA relaying modes in the considered parallel hybrid RF/FSO relay channel.}
\label{Fig:Mode_DU}\vspace{-0.3cm}
\end{figure*}
}

\begin{remk}
In the optimal non-BA protocol, the values of $\alpha_{lm}$, $\beta_{lm}$, and $\rho_l$ depend on the fading states of both the RF and FSO links in the network. In contrast, in the optimal BA protocol, $\alpha_{lm}(\boldsymbol{h}_l)$ is only a function of  the instantaneous CSI of the FSO links and not of the instantaneous CSI of the RF links. Similarly,  $ \beta_{lm}(\boldsymbol{g}_1,\boldsymbol{g}_2)$ and $\rho_l(\boldsymbol{g}_1,\boldsymbol{g}_2)$ are only  functions of  the instantaneous CSI of the RF links and not of the instantaneous CSI of the FSO links. In particular, by comparing the $\Lambda_{1m}^{\mathrm{fso}}({h}_{1m})$ for all the $\mathcal{S}-\mathcal{R}_m$  FSO links, one relay is selected for FSO reception in (\ref{Eq:OptSelec}a),  by comparing the $\Lambda_{2m}^{\mathrm{fso}}({h}_{2m})$ for all the $\mathcal{R}_m-\mathcal{D}$  FSO links, one relay is selected for FSO transmission  in (\ref{Eq:OptSelec}a), and by comparing the $\Lambda_{1m}^{\mathrm{rf}}({g}_{1m})$ and $\Lambda_{2m}^{\mathrm{rf}}({g}_{2m})$ for all the RF links, one relay is selected for either RF reception or RF transmission in (\ref{Eq:OptSelec}b,c). We note that although the optimal $\alpha_{lm}(\boldsymbol{h}_l)$ ($\beta_{lm}(\boldsymbol{g}_1,\boldsymbol{g}_2)$ / $\rho_l(\boldsymbol{g}_1,\boldsymbol{g}_2)$) does not depend on  the instantaneous CSI of the RF (FSO) links,  the statistical CSI of the  RF (FSO) links does affect  $\alpha_{lm}(\boldsymbol{h}_l)$ ($\beta_{lm}(\boldsymbol{g}_1,\boldsymbol{g}_2)$ / $\rho_l(\boldsymbol{g}_1,\boldsymbol{g}_2)$) through Lagrange multiplier $\boldsymbol{\lambda}$, cf. (\ref{Eq:Upd}) and (\ref{Eq:C_Avg}).
\end{remk}
\section{Practical Challenges of the Optimal Protocols}
In this section, we investigate two practical challenges of the optimal protocols, namely the unbounded end-to-end delay of the optimal BA protocol and the global CSI requirement of both the optimal non-BA and BA policies. To cope with these  challenges, we first modify the optimal delay-unconstrained BA policy  given in Theorem~\ref{Theo:Delay-Unlimited} to obtain a delay-constrained BA policy. Subsequently, we present distributed implementations for both the optimal non-BA and BA protocols proposed in Theorem~\ref{Theo:Delay-Limited} and Theorem~\ref{Theo:Delay-Unlimited}, respectively, which require only local CSI knowledge at each node.

\subsection{Delay-Constrained BA Policy}
In the non-BA protocol, the relay nodes are forced to immediately forward the data received from the source to the destination. Therefore, the non-BA protocol is an appropriate option for  applications with stringent delay requirements. On the other hand, in the BA protocol, the relay nodes are allowed to store the data received from the source in their buffers and forward it to the destination when the quality of the relay-destination links is favorable. This leads to  an improvement of the throughput at the expense of  an  increased end-to-end delay. In fact, since there is no limitation on the delay caused by the optimal BA protocol, its end-to-end delay may become unbounded. However, for most practical applications,  it is necessary that the delay remains within a certain range. In the following, we show that a small modification of the optimal BA protocol in Theorem~\ref{Theo:Delay-Unlimited} leads to a delay-constrained protocol whose throughput approaches that of the delay-unconstrained protocol even for small target average delays. 

For the development of the proposed delay-constrained protocol, we limit the size of the buffer at the $m$-th relay to $Q^{\max}_{m},\,\,m\in\{1,\dots,M\}$. Due to the limited buffer size, the  transmit rate of the source to relay $m$ in the $b$-th time slot, denoted by $R_{1m}[b]$, is not only limited by the capacities of the $\mathcal{S}-\mathcal{R}_m$ RF and FSO links, i.e., $C_{1m}^{\mathrm{rf}}[b]$ and $C_{1m}^{\mathrm{fso}}[b]$, but also by the amount of space  available in the buffer of the $m$-th relay, i.e., $Q^{\max}_{m}-Q_m[b-1]$.  Similarly, the rate at which the relay transmits  to the destination in the $b$-th time slot, denoted by $R_{2m}[b]$, is not only limited by the capacities of the $\mathcal{R}_m-\mathcal{D}$ RF and FSO links, i.e., $C_{2m}^{\mathrm{rf}}[b]$ and $C_{2m}^{\mathrm{fso}}[b]$, but also by the amount of   information available in the buffer of the $m$-th relay, i.e., $Q_m[b-1]$. In the following, we present the proposed delay-constrained BA policy.

\textit{Proposed Delay-Constrained BA Policy:} For the  parallel  BA relay channel with hybrid RF/FSO links, the policies for FSO and RF relay selection and optimal RF transmission time allocation given by (\ref{Eq:OptSelec}) in Theorem~\ref{Theo:Delay-Unlimited} lead to a constrained average end-to-end delay if the following modified selection metrics are employed
 \begin{IEEEeqnarray}{rll}\label{Eq:Modified_Metric}
\tilde{\Lambda}_{1m}^{\mathrm{fso}}[b]\,\, &= \lambda_m \min\left\{C_{1m}^{\mathrm{fso}}[b],Q^{\max}_{m}-Q_m[b-1]\right\} \quad \IEEEyesnumber\IEEEyessubnumber \\
\tilde{\Lambda}_{2m}^{\mathrm{fso}}[b]\,\, &= (1-\lambda_m) \min\left\{C_{2m}^{\mathrm{fso}}[b],Q_m[b-1]\right\} \IEEEyessubnumber \\
\tilde{\Lambda}_{1m}^{\mathrm{rf}}[b]\,\, &= \lambda_m \min\left\{C_{1m}^{\mathrm{rf}}[b],Q^{\max}_{m}-Q_m[b-1]\right\} \IEEEyessubnumber \\
\tilde{\Lambda}_{2m}^{\mathrm{rf}}[b]\,\, &= (1-\lambda_m) \min\left\{C_{1m}^{\mathrm{rf}}[b],Q_m[b-1]\right\}, \IEEEyessubnumber 
\end{IEEEeqnarray}
where $\lambda_m,\,\,m\in\{1,\dots,M\}$, is  obtained from (\ref{Eq:Upd}) in Theorem~\ref{Theo:Delay-Unlimited}. Moreover, considering that the optimal values of $\rho_l[b]$ are binary in (\ref{Eq:OptSelec}c), the dynamics of the queue can be simplified with respect to (\ref{Eq:Q_Unlimit}) so that  $Q_m[b]$ is updated in the $b$-th time slot according to
\iftoggle{OneColumn}{%
 \begin{IEEEeqnarray}{rll}\label{Eq:Queue_update}
Q_m[b] = Q_m[b-1] &- \min\left\{\alpha_{2m}[b]C_{2m}^{\mathrm{fso}}[b]+\beta_{2m}[b]\rho_2[b]C_{2m}^{\mathrm{rf}}[b],Q_m[b-1]\right\} \nonumber \\
&+ \min\left\{\alpha_{1m}[b]C_{1m}^{\mathrm{fso}}[b]+\beta_{1m}[b]\rho_1[b]C_{1m}^{\mathrm{rf}}[b],Q^{\max}_{m}-Q_m[b-1]\right\}.
\end{IEEEeqnarray}
}{%
 \begin{IEEEeqnarray}{lll}\label{Eq:Queue_update}
Q_m[b] = Q_m[b-1] \nonumber \\
- \min\left\{\alpha_{2m}[b]C_{2m}^{\mathrm{fso}}[b]+\beta_{2m}[b]\rho_2[b]C_{2m}^{\mathrm{rf}}[b],Q_m[b-1]\right\} \nonumber \\
+ \min\big\{\alpha_{1m}[b]C_{1m}^{\mathrm{fso}}[b]+\beta_{1m}[b]\rho_1[b]C_{1m}^{\mathrm{rf}}[b], \nonumber \\ 
\qquad\qquad Q^{\max}_{m}-Q_m[b-1]\big\}.
\end{IEEEeqnarray}
} 
 Furthermore, the average throughput of the proposed delay-constrained protocol is obtained~as
 \iftoggle{OneColumn}{%
 \begin{IEEEeqnarray}{rll}\label{Eq:Throughput_delay}
\bar{\tau} \,\,& = \sum_m\bar{\tau}_m = \frac{1}{B}\sum_{m}\sum_{b} R_{2m}[b] \nonumber \\
&= \frac{1}{B} \sum_m \sum_{b}\min\left\{\alpha_{2m}[b]C_{2m}^{\mathrm{fso}}[b]+\beta_{2m}[b]\rho_2[b]C_{2m}^{\mathrm{rf}}[b],Q_m[b-1]\right\}.
\end{IEEEeqnarray}
}{%
 \begin{IEEEeqnarray}{rll}\label{Eq:Throughput_delay}
\bar{\tau} \,\,& = \sum_m\bar{\tau}_m = \frac{1}{B}\sum_{m}\sum_{b} R_{2m}[b] \nonumber \\
&= \frac{1}{B} \sum_m \sum_{b}\min\big\{\alpha_{2m}[b]C_{2m}^{\mathrm{fso}}[b] \nonumber \\ 
&\qquad \qquad \qquad +\beta_{2m}[b]\rho_2[b]C_{2m}^{\mathrm{rf}}[b],Q_m[b-1]\big\}. \quad
\end{IEEEeqnarray}
}

 \textit{Average Delay}: The average delay of the proposed protocol is calculated as follows. Let $T[b]$ denote the waiting time (delay) that a bit transmitted by the source in the $b$-th time slot experiences until it reaches the destination. In other words, if a bit is transmitted in the $b$-th time slot by the source and is decoded in the $b'$-th time slot by the destination, the delay for this bit is $T[b]=b'-b$ time slots.  Thereby, according to Little's Law \cite{Little}, the average waiting time/delay of all  data transmitted by the source, denoted by $\bar{T}$, is given by
 \begin{IEEEeqnarray}{rll}\label{Eq:Little}
\bar{T} = \frac{\sum_{m=1}^M \mathbbmss{E}\left\{Q_m[b]\right\}}{\sum_{m=1}^M \mathbbmss{E}\left\{R_{1m}[b]\right\}},
\end{IEEEeqnarray}
where $Q_m[b]$ is given in (\ref{Eq:Queue_update}) and $R_{1m}[b]$ is given by
\iftoggle{OneColumn}{%
 \begin{IEEEeqnarray}{rll}\label{Eq:Input_Rate}
R_{1m}[b] = \min\left\{\alpha_{1m}[b]C_{1m}^{\mathrm{fso}}[b]+\beta_{1m}[b]\rho_1[b]C_{1m}^{\mathrm{rf}}[b],Q^{\max}_{m}-Q_m[b-1]\right\}.
\end{IEEEeqnarray} 
}{%
 \begin{IEEEeqnarray}{rll}\label{Eq:Input_Rate}
R_{1m}[b] = \min\big\{ &\alpha_{1m}[b]C_{1m}^{\mathrm{fso}}[b]+\beta_{1m}[b]\rho_1[b]C_{1m}^{\mathrm{rf}}[b], \nonumber \\
& Q^{\max}_{m}-Q_m[b-1]\big\}.
\end{IEEEeqnarray} 
}

 \begin{remk}
The proposed delay-constrained protocol is able to efficiently limit the average delay by considering not only the instantaneous RF and FSO channel qualities for relay selection and RF time allocation but also the status of the buffers at the relays, cf. (\ref{Eq:Modified_Metric}). Thereby, the smaller the maximum buffer size, i.e., $Q^{\max}_{m}$, the smaller the average delay, i.e., $\bar{T}$. We note that the proposed delay-constrained protocol is \textit{heuristic}. In fact, even for the simple three-node RF relay channel, the optimal policy which maximizes the average throughput for a given average delay is not knwon  \cite{Nikola_B_aided}. However, we show in Section~\ref{Sim} that the proposed heuristic protocol is quite efficient such that for small average delays, e.g. $20$ time slots, it achieves a throughput close to that of the optimal delay-unconstrained protocol in Theorem~\ref{Theo:Delay-Unlimited}.
\end{remk}

\subsection{Distributed Implementation}
The optimal protocols in Theorem~\ref{Theo:Delay-Limited} and Theorem~\ref{Theo:Delay-Unlimited} require global CSI knowledge. On the other hand, relay selection protocols which do not require global CSI knowledge have been proposed in the literature, see e.g. \cite{Survay_BA,Max_Min_Bletsas,Nikola_B_aided}.
In particular, for pure RF communications, the distributed implementation of  relay selection  based on the use of synchronized timers was proposed in \cite{Max_Min_Bletsas} for non-BA relay selection and in \cite{Nikola_B_aided} for BA relay selection. In the following, we present distributed implementations for the non-BA and BA protocols proposed in Theorem~\ref{Theo:Delay-Limited} and Theorem~\ref{Theo:Delay-Unlimited}, respectively.
For distributed implementation of the proposed non-BA and BA protocols, each relay node is required to know only the CSI of the FSO and RF links it is connected to. 

\subsubsection{Distributed Implementation of the Optimal Non-BA Protocol} 

For the optimal non-BA protocol, the proposed distributed implementation involves the following four phases.
 
 \noindent
 \textit{\textbf{Phase I:}})  At the beginning of each time slot, source and destination send pilots to the relay nodes. Then, the relays  estimate the CSI of their respective FSO and RF channels.

 \noindent
 \textit{\textbf{Phase II:}})  To identify the optimal mode, i.e., the hybrid, independent, or mixed mode, each relay has to locally compute the following five throughputs:  the throughput of the hybrid mode, $\tau_{m}^{\mathrm{hyb}}$, using (\ref{Eq:DiffThr}a); the throughput of the involved FSO links, $\tau_{m}^{\mathrm{fso}}$, using (\ref{Eq:DiffThr}b); the throughput of the involved RF links, $\tau_{m}^{\mathrm{rf}}$, using (\ref{Eq:DiffThr}b); and the throughputs $\tau_{m}^{\mathrm{mix1}}= \mathrm{min}\{C_{1m}^{\mathrm{fso}},C_{2m}^{\mathrm{rf}}\}$ and $\tau_{m}^{\mathrm{mix2}} = \mathrm{min}\{C_{1m}^{\mathrm{rf}},C_{2m}^{\mathrm{fso}}\}$. Note that these five throughputs can be calculated at each relay node  based on the CSI of the FSO and RF links to which it is directly connected.

 \noindent
 \textit{\textbf{Phase III:}})  Each relay sets five timers $T_m^{\mathrm{hyb}}$, $T_m^{\mathrm{fso}}$, $T_m^{\mathrm{rf}}$, $T_{m}^{\mathrm{mix1}}$, and $T_{m}^{\mathrm{mix2}}$ which expire after $\eta/\tau_{m}^{\mathrm{hyb}}$, $\eta/\tau_{m}^{\mathrm{fso}}$, $\eta/\tau_{m}^{\mathrm{rf}}$, $\eta/\tau_{m}^{\mathrm{mix1}}$, and $\eta/\tau_{m}^{\mathrm{mix2}}$ seconds, respectively, where $\eta$ is a constant which scales the expiry time into a reasonable range. For each $\zeta \in \{\mathrm{hyb}, \mathrm{fso}, \mathrm{rf}, \mathrm{mix1}, \mathrm{mix2}\}$, the relay whose timer $T_m^{\mathrm{\zeta}}$ expires first broadcasts beacon $B_m^{\zeta}$ which contains the information of the relay index $m$ and the timer index $\zeta$. At the same time, all relay nodes listen and if they receive beacon $B_m^{\mathrm{\zeta}},\,\,\zeta \in \{\mathrm{hyb}, \mathrm{fso}, \mathrm{rf}, \mathrm{mix1}, \mathrm{mix2}\},$ from another relay, they do not emit their own beacon  $B_m^{\mathrm{\zeta}}$.
 
 \noindent
 \textit{\textbf{Phase IV:}})  After transmission of the beacons, all the nodes decode the information of each transmitted beacon and determine the index of the relays with maximum $\tau_{m}^{\mathrm{\zeta}},\,\forall \mathrm{\zeta}$. Moreover, by measuring the expiry time of the timers which expired first, all the nodes can calculate the corresponding maximum throughput for each $\mathrm{\zeta}$ as $\underset{m}{\mathrm{max}}\,\,\tau_{m}^{\mathrm{\zeta}}=\eta/T_{m}^{\mathrm{\zeta}}$. 
   Hence, the nodes 
  are able to calculate the maximum throughputs of the hybrid mode, $\underset{m}{\mathrm{max}}\,\,\tau_{m}^{\mathrm{hyb}}$, the independent mode, $\underset{m}{\mathrm{max}}\,\, \tau_{m}^{\mathrm{fso}}+\underset{m}{\mathrm{max}}\,\, \tau_{m}^{\mathrm{rf}}$, and the mixed mode, $\underset{m}{\mathrm{max}}\,\, \tau_{m}^{\mathrm{mix1}}+\underset{m}{\mathrm{max}}\,\, \tau_{m}^{\mathrm{mix2}}$, and  can distributedly determine the optimal mode as the one with the maximum throughput among the candidate hybrid,  independent, and mixed modes and the corresponding optimal RF/FSO relays.

\subsubsection{Distributed Implementation of the Optimal BA Protocol} 
The proposed distributed implementation of the optimal BA protocol involves four phases as follows.

 \noindent
 \textit{\textbf{Phase I:}})   At the beginning of each time slot, source and destination transmit pilots to the relay nodes. Then, the relays estimate the CSI of their respective FSO and RF channels. 

 \noindent
 \textit{\textbf{Phase II:}})  To select the best relays, each relay has to compute its respective selection metrics given in Theorem~2, i.e., 
 $\Lambda_{lm}^{\mathrm{fso}}({h}_{lm})$ and $\Lambda_{lm}^{\mathrm{rf}}({g}_{lm}),\,\,l=1,2,$  as follows. Each relay calculates the capacities of its respective FSO and RF links, i.e., $C_{lm}^{\mathrm{fso}}[b]$ and $C_{lm}^{\mathrm{rf}}[b],\,\,l=1,2,$ using (\ref{Eq:FSO_C}) and (\ref{Eq:RF_C}), respectively. Moreover, $\lambda_m$ is a constant and can be obtained offline at the beginning of the transmission process using (\ref{Eq:Upd}). Using $\lambda_m$ and the capacities of the involved RF and FSO links, each relay is able to calculate its respective selection metrics in each time slot.

 \noindent
 \textit{\textbf{Phase III:}})  Each relay sets three timers $T_{m}^{\mathrm{fso1}}$, $T_{m}^{\mathrm{fso2}}$, and $T_{m}^{\mathrm{rf}}$ which expire after $\eta/\Lambda_{1m}^{\mathrm{fso}}({h}_{1m})$, $\eta/\Lambda_{2m}^{\mathrm{fso}}({h}_{2m})$, and 
 $\eta/\mathrm{max}\lbrace\Lambda_{1m}^{\mathrm{rf}}({g}_{1m}),\Lambda_{2m}^{\mathrm{rf}}({g}_{2m})\rbrace$ seconds, respectively. For the FSO links, for each $\xi\in\{\mathrm{fso1},\mathrm{fso2}\}$, the relay whose timer $T_{m}^{\xi}$ expires first broadcasts beacon $B_m^{\xi}$. For the RF links, the relay whose timer $T_{m}^{\mathrm{rf}}$ expires first broadcasts beacon $B_m^{\mathrm{rf1}}$ if $\mathrm{max}\lbrace\Lambda_{1m}^{\mathrm{rf}}({g}_{1m}),\Lambda_{2m}^{\mathrm{rf}}({g}_{2m})\rbrace=\Lambda_{1m}^{\mathrm{rf}}({g}_{1m})$  and beacon $B_m^{\mathrm{rf2}}$ if $\mathrm{max}\lbrace\Lambda_{1m}^{\mathrm{rf}}({g}_{1m}),\Lambda_{2m}^{\mathrm{rf}}({g}_{2m})\rbrace$ $=\Lambda_{2m}^{\mathrm{rf}}({g}_{2m})$. The beacons contain  information about the relay index $m$ and whether the relay is selected for RF/FSO reception or RF/FSO transmission.

 \noindent
 \textit{\textbf{Phase IV:}})  The nodes which transmit beacons  are the selected relays. Hence, after transmission and reception of the beacons, each node knows which relays are selected for RF/FSO reception and transmission. 

\section{Simulation Results}\label{Sim}
In this section, we first present  benchmark schemes for the proposed relay selection policies. Subsequently, we evaluate the performances of the proposed protocols and compare them with those of the benchmark schemes.

\subsection{Benchmark Schemes}

As benchmark scheme for the non-BA protocol, we consider the well-known max-min relay selection protocol \cite{Max_Min_Bletsas,Relay_Selection_Schober} where for each fading state, the relay with the maximum bottleneck capacity, i.e., the minimum of the capacities of the $\mathcal{S}-\mathcal{R}_m$ and $\mathcal{R}_m-\mathcal{D}$ links, is selected. A recent overview of  BA relay selection protocols is provided in  \cite{Survay_BA}. For the BA case, we select the scheme in \cite{Nikola_B_aided} as benchmark scheme for the proposed BA protocol where, in each  time slot, the optimal relay is selected such that the end-to-end throughput is maximized. We note that the  protocol in  \cite{Nikola_B_aided} outperforms  the  other BA protocols  available in the literature including the max-max protocol in \cite{Max_Max} and the max-link protocol in \cite{Max_link} in terms of the achievable rate. 

 More in detail, we employ  the protocols in  \cite{Nikola_B_aided} and \cite{Max_Min_Bletsas} for the following two scenarios: \textit{i) FSO only:}  Relay selection for the FSO links without RF links as backups \cite{FSO_Diversity_Chadi, Relay_Selection_Schober} and \textit{ii) Independent RF/FSO relay selection:}  Relay selection  and data transmission are performed independently for the RF and FSO links. In the non-BA benchmark schemes, we assume that for the  RF links, each time slot is divided into two sub-time slots of equal length for  $\mathcal{S}-\mathcal{R}_m$ and $\mathcal{R}_m-\mathcal{D}$ RF transmission. We compare the proposed protocols with the FSO-only protocols  to quantify the performance gain introduced by  RF backup links. Moreover, we consider the independent RF/FSO protocol to evaluate the benefits of the proposed optimal transmission strategies in hybrid RF/FSO systems.

\begin{table}
\label{Table:Parameter}
\caption{Default Values for System Parameters \cite{HybridRFFSOShi,Schober}.\vspace{-0.2cm}} 
\begin{center}
\scalebox{0.7} { 
\begin{tabular}{|| c | c || c | c ||}
  \hline
   \multicolumn{2}{||c||}{\textbf{RF Link}} & \multicolumn{2}{|c||}{\textbf{FSO Link}} \\ \hline \hline    
 Parameter &  Value & Parameter &  Value \\ \hline \hline 
 $G^{\mathrm{rf}}_{\mathrm{tx}},G^{\mathrm{rf}}_{\mathrm{rx}}$ & $10$ dBi & $R$ & $0.5\frac{1}{\mathrm{V}}$  \\  \hline
 $P_{\mathcal{S}}^{\mathrm{rf}},P_{\mathcal{R}}^{\mathrm{rf}}$ & $0.2$ mW ($23$ dBm) & $P_{\mathcal{S}}^{\mathrm{fso}},P_{\mathcal{R}}^{\mathrm{fso}}$ & $20$ mW ($13$ dBm) \\  \hline
  $N_0$ & $-114$ dBm/MHz & $\sigma^2$ &  $10^{-14}$ $\mathrm{A}^2$\\ \hline 
   $\lambda^{\mathrm{rf}}$ &  $85.7$ mm ($3.5$ GHz) & $\lambda^{\mathrm{fso}}$ & $1550$ nm ($193$ THz)\\ \hline 
      $W^{\mathrm{rf}}$ &  $20$ MHz & $W^{\mathrm{fso}}$ & $1$ GHz  \\ \hline 
    $(\Omega,\Psi)$ & $(4,1)$  & $(\Theta,\Phi)$ & $(2.23,1.54)$ \\  \hline
    $\nu$ & $3.5$ & $k$ & $0.032$ (light-moderate fog)\\ \hline 
      $N_F$ &  $5$ dB &$r$ &   $10$ cm \\ \hline 
 $d^{\mathrm{rf}}_{\mathrm{ref}}$ & $80$ m & $\phi$   & $2$ mrad \\ \hline         
\end{tabular}
}
\end{center}
\vspace{-0.7cm}
\end{table}

\subsection{Performance Evaluation}
Unless otherwise stated, the values of the parameters for the RF and FSO links used  to produce  the  simulation  results  reported in  this  section are given in Table~I. In particular, we generated random fading realizations for $B=10^5$ time slots, applied the proposed and the benchmark relay selection policies in each time slot, and computed the throughput for each policy  as the average data rate received at the destination using that policy. 

\begin{figure*}[!tbp]
  \centering
  \begin{minipage}[b]{0.47\textwidth}
  \centering
\resizebox{1\linewidth}{!}{\psfragfig{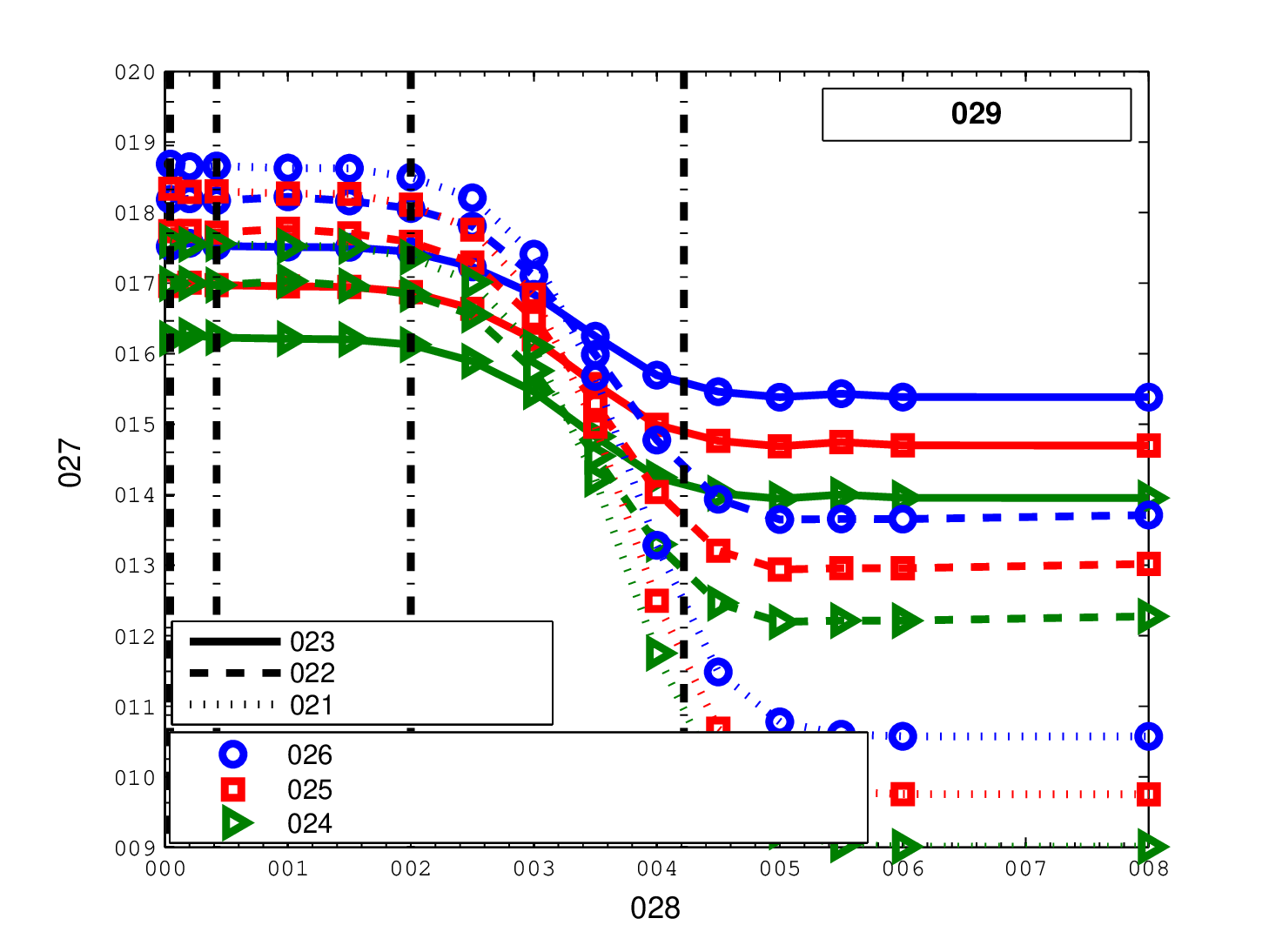} }  
\caption{Average throughput, $\bar{\tau}$, in Mbits/second vs. FSO weather-dependent attenuation factor, $k\times 10^{-3}$, for $M=3$ and $d_{1m}=d_{2m}=800$ m (non-BA case).
 From low  to  high  values  of  $k$,  the  vertical  dashed-dotted  lines  represent  the
following weather conditions \cite{Schober}: clear air, haze, light fog, and moderate fog,  respectively.}
\label{Fig:Thr_Kappa_NonBuffer}
  \end{minipage}
    \hfill
  \begin{minipage}[b]{0.1\textwidth}
  \end{minipage}
  \hfill
  \begin{minipage}[b]{0.47\textwidth}
  \centering
\resizebox{1\linewidth}{!}{\psfragfig{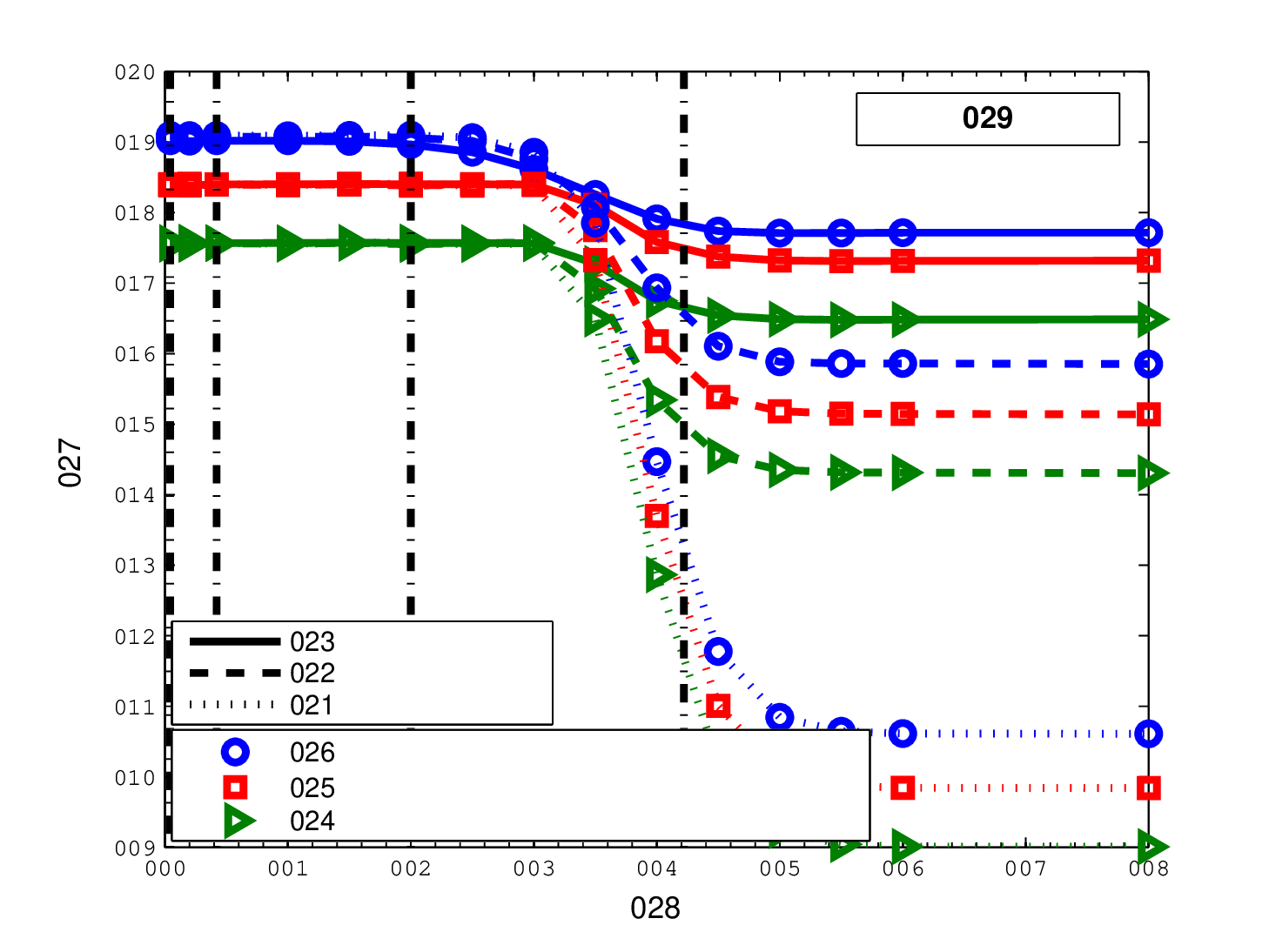} }  
\caption{Average throughput, $\bar{\tau}$, in Mbits/second vs. FSO weather-dependent attenuation factor, $k\times 10^{-3}$, for $M=3$ and $d_{1m}=d_{2m}=800$ m (BA case).
 From low  to  high  values  of  $k$,  the  vertical  dashed-dotted  lines  represent  the
following weather conditions \cite{Schober}: clear air, haze, light fog, and moderate fog, respectively.}
\label{Fig:Thr_Kappa_Buffer}
  \end{minipage}
    \hfill
  \begin{minipage}[b]{0.02\textwidth}
  \end{minipage}\vspace{-0.4cm}
\end{figure*}

In Figs.~\ref{Fig:Thr_Kappa_NonBuffer} and \ref{Fig:Thr_Kappa_Buffer}, we show the average throughput vs. the weather-dependent  attenuation factor of the FSO links, $k$, for the non-BA and BA protocols, respectively.  We assume $M=3$, $d_{1m}=d_{2m}=800$ m, and consider the  following three scenarios. In the first scenario, we vary only  $k_{11}=k$;  in the second scenario, we vary  $k_{11}=k_{12}=k$; and in the third scenario, we vary  $k_{11}=k_{12}=k_{13}=k$, i.e., the weather-dependent attenuation factors of all FSO links in the first hop. The considered scenarios reflect the fact that different FSO links may be affected by different weather conditions, e.g. passing clouds or birds may affect only some of the FSO links.  From Figs.~\ref{Fig:Thr_Kappa_NonBuffer} and \ref{Fig:Thr_Kappa_Buffer}, we observe that the throughput decreases as  $k$ increases. Moreover, as $k\to \infty$, all  throughputs saturate at certain values representing the case where the corresponding FSO links are not available anymore. For instance, for the FSO-only protocol in the third scenario, the throughput drops to zero as $k\to \infty$  since  all the FSO links of the first hop become unavailable. In contrast, the proposed protocol achieves a  non-zero throughput because of the RF back-up links and outperforms the independent RF/FSO protocol. Furthermore, by comparing the curves in  Figs.~\ref{Fig:Thr_Kappa_NonBuffer} and \ref{Fig:Thr_Kappa_Buffer}, we observe that the BA protocols achieve higher throughputs than the corresponding non-BA protocols.

\begin{figure*}[!tbp]
  \centering
  \begin{minipage}[b]{0.47\textwidth}
  \centering
\resizebox{1\linewidth}{!}{\psfragfig{} }  
\caption{Average throughput, $\bar{\tau}$, in Mbits/second vs. RF transmit power, $P_{\mathcal{S}}^{\mathrm{rf}}=P_{\mathcal{R}}^{\mathrm{rf}} = P^{\mathrm{rf}}$, for $M=3$, $d_{1m}=1000$ m, and $d_{2m}=800$ m.}
\label{Fig:Thr_SNR_RF_Asym}
  \end{minipage}
    \hfill
  \begin{minipage}[b]{0.1\textwidth}
  \end{minipage}
  \hfill
  \begin{minipage}[b]{0.47\textwidth}
  \centering
\resizebox{1\linewidth}{!}{\psfragfig{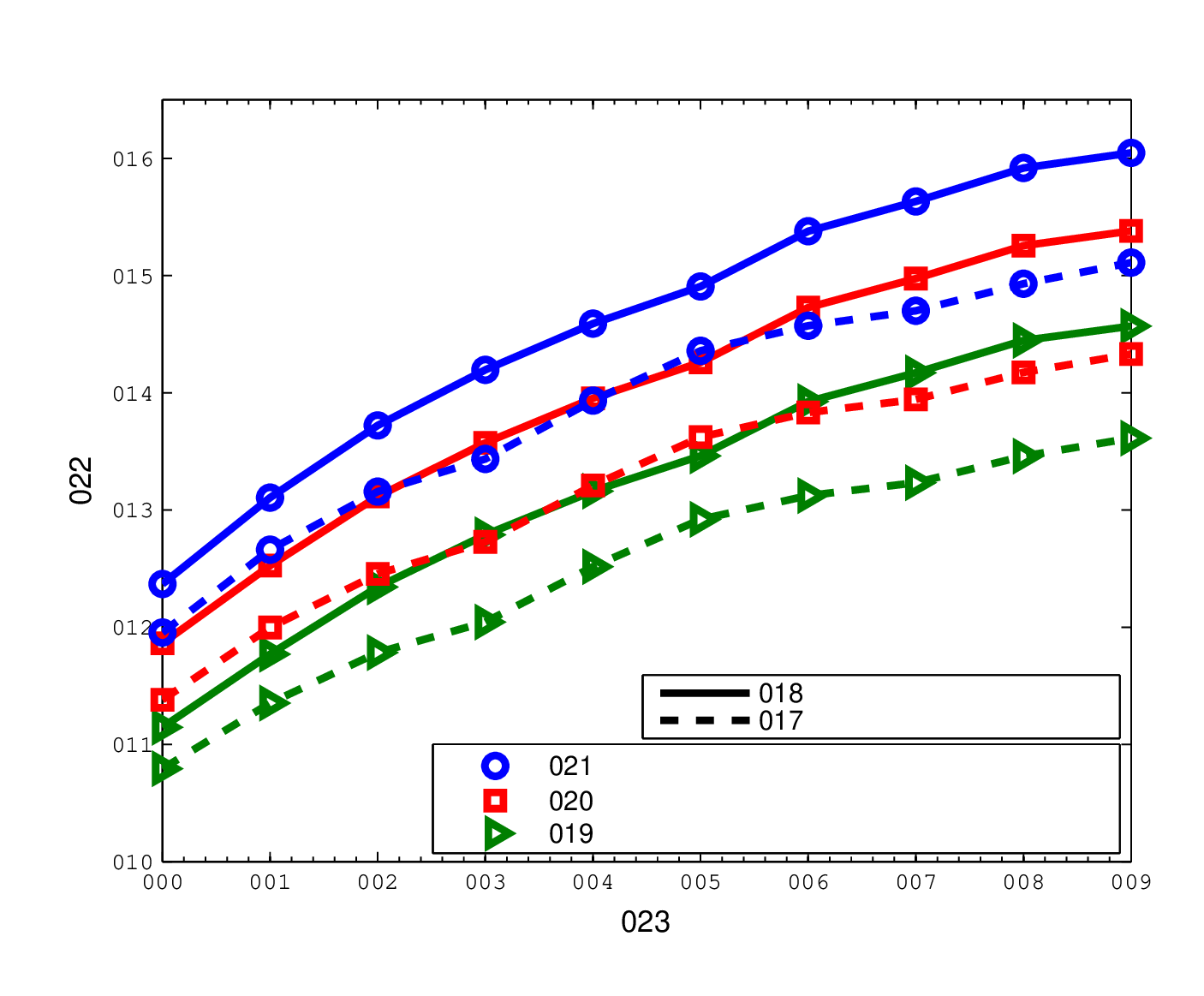} }
\caption{Average throughput, $\bar{\tau}$, in Mbits/second vs. number of relays, $M$, for $d_{1m}=1000$ m and $d_{2m}=800$ m.}
\label{Fig:Thr_M_Asym}
  \end{minipage}
    \hfill
  \begin{minipage}[b]{0.02\textwidth}
  \end{minipage}\vspace{-0.4cm}
\end{figure*}

In Fig.~\ref{Fig:Thr_SNR_RF_Asym}, the average throughput vs. the RF transmit power is shown for $M=3$, $d_{1m}=1000$~m, and $d_{2m}=800$~m for both non-BA and BA relays. As can be seen from Fig.~\ref{Fig:Thr_SNR_RF_Asym}, the average throughputs of the independent RF/FSO protocols and the proposed protocols increase with increasing  RF transmit power whereas the throughputs of the FSO-only protocols do not depend on the RF transmit power. Moreover, due to optimal joint relay selection for the RF and FSO links, the proposed protocols not only outperform the independent RF/FSO protocols  for both the non-BA and the BA cases but also achieve a higher multiplexing gain for the considered range of RF transmit powers. Furthermore, as expected, the BA protocols considerably outperform the non-BA protocols.

 In Fig.~\ref{Fig:Thr_M_Asym}, we show the average throughput vs. the number of  relay nodes for $d_{1m}=1000$ m and $d_{2m}=800$ m for both non-BA and BA relays. From this figure, we observe that by increasing the number of relays, the throughput can be considerably improved due to the available spatial diversity. For instance, for the proposed BA protocol, we observe throughput improvements of $95\%$ and $150\%$ for $M=5$ and $M=10$, respectively, compared to the case of $M=1$. Fig.~\ref{Fig:Thr_M_Asym} also confirms that the proposed protocols outperform all considered benchmark schemes by a large margin.

\begin{figure*}[!tbp]
  \centering
  \begin{minipage}[b]{0.47\textwidth}
  \centering
\resizebox{1\linewidth}{!}{\psfragfig{} }  
\caption{Average throughput, $\bar{\tau}$, in Mbits/second vs. average delay, $\bar{T}$, for $M\in\{1,3,5\}$  and  $d_{1m}=d_{2m}=800$~m.}
\label{Fig:Thr_Delay_M}
  \end{minipage}
    \hfill
  \begin{minipage}[b]{0.1\textwidth}
  \end{minipage}
  \hfill
  \begin{minipage}[b]{0.47\textwidth}
  \centering
\resizebox{1\linewidth}{!}{\psfragfig{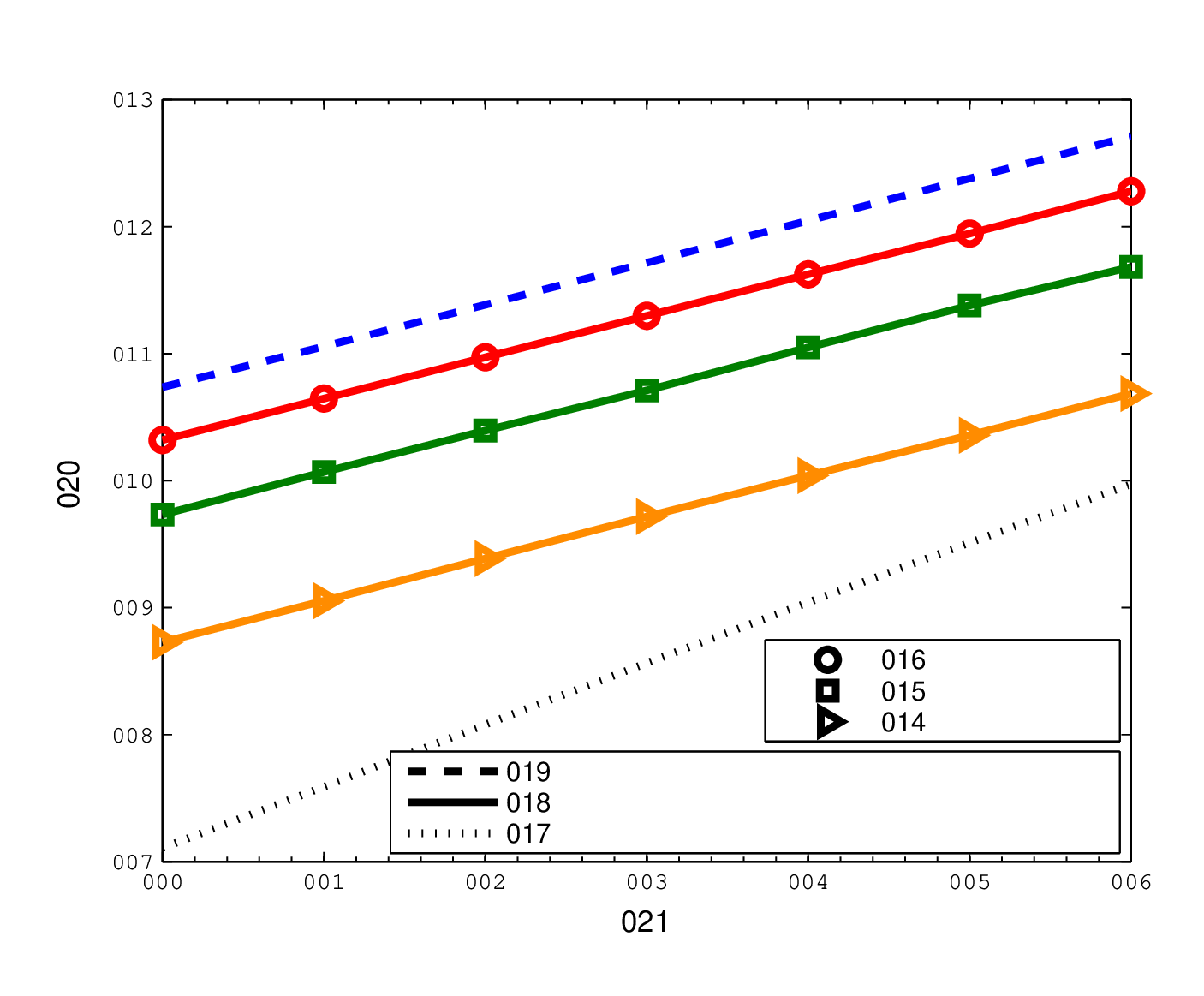} }  
\caption{Average throughput, $\bar{\tau}$, in Mbits/second vs. RF transmit power, $P_{\mathcal{S}}^{\mathrm{rf}}=P_{\mathcal{R}}^{\mathrm{rf}} = P^{\mathrm{rf}}$, for $M=3$ and $d_{1m}=d_{2m}=800$ m.}
\label{Fig:Thr_Delay_Block}
  \end{minipage}
    \hfill
  \begin{minipage}[b]{0.02\textwidth}
  \end{minipage}\vspace{-0.4cm}
\end{figure*}

Recall that the gains that the BA protocols achieve compared to the non-BA protocols come at the expense of  an unbounded end-to-end delay. Therefore, in Fig.~\ref{Fig:Thr_Delay_M}, we study the performance of the delay-constrained BA protocol developed in Subsection~IV.A. In particular, in Fig.~\ref{Fig:Thr_Delay_M}, we show the average throughput vs. the average delay  for $M \in\{1,3,5\}$ and  $d_{1m}=d_{2m}=800$~m. For each point on the curves for the proposed delay-constrained BA protocol, we chose an appropriate value for $Q^{\mathrm{max}}$ which led to the desired delay. Additionally, Fig.~\ref{Fig:Thr_Delay_M} includes results for the non-BA and the delay-unconstrained BA protocols as  lower and upper bounds for the throughput with average delays of $\bar{T}\leq 1$ and $\bar{T}\to\infty$ time slots, respectively. We observe that for sufficiently large target average delays, the throughput of the delay-constrained protocol approaches the delay-unconstrained upper bound which reveals the effectiveness of the proposed delay-constrained protocol. 

To further investigate the performance of the proposed delay-constrained protocol,  in Fig.~\ref{Fig:Thr_Delay_Block}, we plot the average throughput vs. the RF transmit power  for $M=3$ and   $d_{1m}=d_{2m}=800$ m for delays of $\bar{T} \in \{5,10,20\}$ time slots. Fig.~\ref{Fig:Thr_Delay_Block} reveals that as the allowed delay  increases, the achievable throughput improves. Furthermore, for a delay of $20$ time slots, the proposed delay-constrained protocol significantly outperforms the non-BA protocol and achieves an average throughput  close to the upper bound for  the considered range of RF transmit powers. 

\section{Conclusions}
We investigated the problem of throughput maximization for the parallel hybrid RF/FSO relay channel.  Thereby, we distinguished two  cases depending on whether  or not the relays are equipped with buffers. For both cases, we derived the optimal relay selection policies for transmission and reception for the RF and FSO links and the optimal time allocation policies for  RF transmission and reception. Additionally, since the optimal BA policy introduces unbounded delay, we proposed a delay-constrained BA policy which ensures a certain target average end-to-end delay. Furthermore, we developed distributed implementations of the proposed optimal non-BA and BA policies. Simulation results verified the superiority of the proposed adaptive protocols compared to benchmark schemes from the literature, especially when the FSO links suffered from severe atmospheric impairments. Furthermore, even for an average delay of only $20$ time slots, the proposed delay-constrained BA  protocol considerably outperformed the optimal non-BA protocol and achieved a performance close to that of the optimal delay-unconstrained BA protocol.

\appendices

\section{Proof of Theorem \ref{Theo:Delay-Limited}}\label{App:Theo2}
In this appendix, we derive the solution to the optimization problem in (\ref{Eq:Opt_DL}). 
To this end, we first specify the potential candidates for the optimal relay selection policy among all  possible relay selection policies $(\alpha_{lm},\beta_{lm})$. Subsequently, we derive the  optimal RF time allocation policy $\rho_l^*$ for each of the potential candidates for the optimal relay selection policy. Finally, the  relay selection policy  which yields the maximum end-to-end throughput among the candidate relay selection policies is chosen as the optimal relay selection policy $(\alpha_{lm}^*,\beta_{lm}^*)$. 

\subsection{Candidate Policies}
  
The feasible sets $\boldsymbol{\mathcal{A}}$ and $\boldsymbol{\mathcal{B}}$ of the relay selection variables
allow the selection of at most four different relays for RF/FSO reception and transmission. Therefore, there are in total $M^4$ possibilities for the optimal binary values of $\alpha_{lm}$ and $\beta_{lm}$ in the feasible set $\boldsymbol{\mathcal{A}}\times\boldsymbol{\mathcal{B}}$. However, due to the constraints in (\ref{Eq:Opt_DL}), the optimal relay selection policy can select at most two different relays for RF/FSO reception and transmission in order to  ensure  that  the  data  which  is  transmitted  from  the
source to a certain relay can be actually forwarded to the destination. Thereby, there are $\frac{M(M-1)}{2} $ possibilities to select two relays out of $M$ relays. Moreover, for a given selected relay pair, there are $2^4=16$ possibilities to assign the selected relays to RF/FSO reception and transmission, respectively. In the following, we show that only $6$ among these $16$ possibilities are candidates for the optimal relay selection policy. To this end, let $m$ and $n$ be the indices of the selected relays. Considering the feasible sets $\boldsymbol{\mathcal{A}}$ and $\boldsymbol{\mathcal{B}}$, we investigate the following $2^2=4$ possibilities for the RF/FSO receiving relays: \textit{i)} Relay $m$ is selected for both RF/FSO reception, i.e., $\alpha_{1m}=\beta_{1m}=1$. In this case, relay $m$ is the only option for RF/FSO transmission, i.e., $\alpha_{2m}=\beta_{2m}=1$ has to hold (hybrid mode). \textit{ii)} Relay $m$ is selected for RF reception and relay $n$ is selected for FSO reception, i.e., $\alpha_{1n}=\beta_{1m}=1$. Here, there are two options, namely, relays $m$ and $n$ are chosen either for RF and FSO transmission, respectively, i.e., $\alpha_{2n}=\beta_{2m}=1$  (independent mode), or for FSO and RF transmission, respectively, i.e., $\alpha_{2m}=\beta_{2n}=1$  (mixed mode). Cases \textit{iii)} and \textit{iv)} are identical to Cases \textit{i)} and \textit{ii)}, respectively, after changing the roles of relays $n$ and $m$. To summarize, among the $M^4$ possibilities for $\alpha_{lm}$ and $\beta_{lm}$ in the feasible set $\boldsymbol{\mathcal{A}}\times\boldsymbol{\mathcal{B}}$, only $3M(M-1)$ possibilities have to be investigated for the optimal relay selection policy. 

\subsection{Optimal RF Time Allocation }
 
In the following, the  optimal RF time allocation policy $\rho_l^*$ and the resulting throughput are derived for  the aforementioned $3M(M-1)$ possibilities depending on their modes of transmission, namely the hybrid, independent, and mixed modes.

\noindent \textit{Case 1 (Hybrid Mode):}   Suppose relay $\mathcal{R}_m$ is selected for both RF/FSO transmission/reception. Thereby, the optimal $\rho_l$ is found such that the minimum of the capacities of the  $\mathcal{S}-\mathcal{R}_m$ hybrid RF/FSO link and the $\mathcal{R}_m-\mathcal{D}$ hybrid RF/FSO link is maximized, i.e.,
\begin{IEEEeqnarray}{lll}\label{Eq:DL_Rho_Backup}
\rho_1 = 1 - \rho_2 = \left[ \frac{C_{2m}^{\mathrm{fso}}+C_{2m}^{\mathrm{rf}}-C_{1m}^{\mathrm{fso}}}{C_{1m}^{\mathrm{rf}}+C_{2m}^{\mathrm{rf}}} \right]_0^1,
\end{IEEEeqnarray}
which leads to the overall throughput  $\tau_m^{\mathrm{hyb}}$ given in (\ref{Eq:DiffThr}a). 
Moreover, the optimal relay for RF and FSO
transmission is the one which leads to the maximum value of $\tau_m^{\mathrm{hyb}}$ in (\ref{Eq:DiffThr}a), i.e., the index of the optimal relay  is given by $m^* = \underset{m}{\mathrm{argmax}}\,\, \tau_{m}^{\mathrm{hyb}}$.

\noindent \textit{Case 2 (Independent Mode):} Let relay $\mathcal{R}_m$ be selected for both FSO reception and transmission and a different relay $\mathcal{R}_n$  be selected for RF reception and transmission. The optimal $\rho_l$ which makes the
RF transmission rates of the $\mathcal{S}-\mathcal{R}_n$ and $\mathcal{R}_n-\mathcal{D}$ links equal is found as
\begin{IEEEeqnarray}{lll}\label{Eq:DL_Rho_Indep}
\rho_1 = 1 - \rho_2=   \frac{C_{2n}^{\mathrm{rf}}}{C_{1n}^{\mathrm{rf}}+C_{2n}^{\mathrm{rf}}}.
\end{IEEEeqnarray}
This leads to the overall throughput  $\tau_{mn}^\mathrm{ind}$ given in (\ref{Eq:DiffThr}b). Moreover, in this case, we can independently select the relay which maximizes the throughput of FSO transmission, i.e., $m^* = \underset{m}{\mathrm{argmax}}\,\, \tau_{m}^{\mathrm{fso}}$, and the relay which maximizes the throughput of RF transmission, i.e., $n^*=\underset{n}{\mathrm{argmax}}\,\, \tau_{n}^{\mathrm{rf}}$.

\noindent \textit{Case 3 (Mixed Mode):} Here, different relays $\mathcal{R}_m$ and $\mathcal{R}_n$ are selected for FSO reception and transmission, respectively. Moreover, for this case to be optimal,  relays $\mathcal{R}_m$ and $\mathcal{R}_n$ have to be selected for RF transmission and RF reception, respectively. For this case, we can distinguish the following four subcases depending on which links are the bottleneck for data transmission.
 
\textit{Subcase 1:} The bottleneck links for both relays $\mathcal{R}_m$ and $\mathcal{R}_n$ are the FSO links. Hence, the RF time sharing variables have to be chosen to support the FSO links, i.e., $\rho_1 \geq   \frac{C_{2n}^{\mathrm{fso}}}{C_{1n}^{\mathrm{rf}}}$ and  $\rho_2 \geq   \frac{C_{1m}^{\mathrm{fso}}}{C_{2m}^{\mathrm{rf}}}$. Therefore, a necessary condition for this subcase to be optimal is that $\frac{C_{2n}^{\mathrm{fso}}}{C_{1n}^{\mathrm{rf}}}+\frac{C_{1m}^{\mathrm{fso}}}{C_{2m}^{\mathrm{rf}}}\leq 1$ holds. Without loss of generality and since $\rho_1+\rho_1=1$ has to hold, we choose the following solution 
\begin{IEEEeqnarray}{lll}\label{Eq:DL_Rho_Cross1}
\rho_1 = 1 - \rho_2 =   \frac{C_{2n}^{\mathrm{fso}}}{C_{1n}^{\mathrm{rf}}}.
\end{IEEEeqnarray}
This subcase leads to throughput $\tau = C_{1m}^{\mathrm{fso}}+C_{2n}^{\mathrm{fso}}$. 

\textit{Subcase 2:} The bottleneck links for both relays $\mathcal{R}_m$ and $\mathcal{R}_n$ are the RF links. This leads to throughput $\tau = \rho_1 C_{1n}^{\mathrm{rf}}+ \rho_2C_{2m}^{\mathrm{rf}}$. Hence, we obtain 
\begin{IEEEeqnarray}{lll}\label{Eq:DL_Rho_Cross1}
\rho_1 = 1 - \rho_2 =   \begin{cases}
1,\quad &\mathrm{if} \,\, C_{1n}^{\mathrm{rf}} \geq C_{2m}^{\mathrm{rf}}\\
0, &\mathrm{otherwise}. 
\end{cases}
\end{IEEEeqnarray}
However, the RF transmission time allocation policy in (\ref{Eq:DL_Rho_Cross1}) implies that the RF link is selected to support either  FSO transmission or reception, i.e., only one of the relays is active. Therefore, this subcase cannot be optimal since Case 1 always yields a higher throughput.

\textit{Subcase 3:} The bottleneck links for relays $\mathcal{R}_m$ and $\mathcal{R}_n$ are the FSO and RF links, respectively.  This leads to throughput $\tau  = C_{1m}^{\mathrm{fso}}+ \rho_1C_{1n}^{\mathrm{rf}}$. Here, the throughput can be always improved by increasing  $\rho_1$ and decreasing $\rho_2$ until the $\mathcal{S}-\mathcal{R}_m$ FSO link is no longer the bottleneck. This contradicts the earlier assumption of this subcase, i.e., Subcase 3 cannot occur for the optimal solution.

\textit{Subcase 4:} The bottleneck links for relays $\mathcal{R}_m$ and $\mathcal{R}_n$ are the RF and FSO links, respectively. Similar to Subcase 3, Subcase 4 cannot occur for the optimal solution.

To conclude, among the four possible subcases for Case 3, only Subcase 1 can be the optimal solution for some fading realizations. Hence, without loss of generality, we define the throughput of Case 3, denoted by $\tau_{mn}^{\mathrm{mix}}$, in (\ref{Eq:DiffThr}c) as the throughput of Subcase 1 if the necessary condition for this subcase, i.e., $\frac{C_{2n}^{\mathrm{fso}}}{C_{1n}^{\mathrm{rf}}}+\frac{C_{1m}^{\mathrm{fso}}}{C_{2m}^{\mathrm{rf}}}\leq 1$ holds, and zero otherwise. The indices of the optimal relays are given by $(m^*,n^*) = \underset{(m,n)}{\mathrm{argmax}}\,\, \tau_{mn}^{\mathrm{mix}}$.

\subsection{Optimal Policy} 
 
Now, the remaining question is in which mode the RF and FSO links should operate for a given channel realization.
Since our goal is to maximize the throughput, we have to select the case which yields the maximum achievable throughput, i.e., the maximum value among $\tau_{m^*}^{\mathrm{hyb}}$, $\tau_{m^*}^{\mathrm{fso}}+\tau_{n^*}^{\mathrm{rf}}$, and $\tau_{m^*n^*}^{\mathrm{mix}}$. This leads to the relay selection policy given in Theorem 1 and completes the proof.

\section{Proof of Theorem \ref{Theo:Delay-Unlimited}}\label{App:Theo}
In this appendix, our aim is to solve the optimization problem in (\ref{Eq:Opt_DUL}). The   problem  in (\ref{Eq:Opt_DUL}) is non-convex because of the binary constraints on $\alpha_{lm}[b]$ and $\beta_{lm}[b]$ and the multiplication of two variables, $\beta_{lm}[b] \rho_l[b]$. To make the problem tractable,  we  relax the binary constraint $\alpha_{lm}[b] \in \lbrace 0,1\rbrace$ to $\alpha_{lm}[b] \in [0,1]$ and define new variable $\gamma_{lm}[b] \triangleq \beta_{lm}[b] \rho_l[b]$. The  feasible sets of the new variables of the relaxed problem are given by $\alpha_{lm}[b] \in\widetilde{\boldsymbol{\mathcal{A}}}=\{\boldsymbol{\alpha}|\alpha_{lm}[b]  \in [0,1],\,\, \forall l,m,b\,\,\wedge\,\,\sum_{\forall m}\alpha_{lm}[b]=1,\,\,\forall l,b \}$ and $\gamma_{lm}[b] \in\boldsymbol{\mathcal{G}}=\{\boldsymbol{\gamma}|\gamma_{lm}[b]  \in [0,1],\,\, \forall l,m,b\,\,\wedge\,\,\sum_{\forall l} \sum_{\forall m}\gamma_{lm}[b]=1,\,\,\forall b \}$ where $\boldsymbol{\gamma}$ is a vector containing the $\gamma_{lm}[b]$, $\forall l,m,b$. The relaxed problem is linear and can be solved globally using the dual Lagrange  method \cite{Boyd}. Moreover, we will show that the solution of the relaxed problem is binary, and hence, also solves  the original problem in (\ref{Eq:Opt_DUL}). In particular, the Lagrangian function corresponding to the relaxed version of the optimization problem in (\ref{Eq:Opt_DUL}) is obtained as
\iftoggle{OneColumn}{%
\begin{IEEEeqnarray}{cll}\label{Eq:Lag}
\mathcal{L}(\boldsymbol{\bar{\tau}},\boldsymbol{\alpha},\boldsymbol{\gamma},\boldsymbol{\bar{\lambda}})=&\sum_{\forall m}\bar{\tau}_m+\sum_{\forall m}\lambda_{1m}\left( \dfrac{1}{B} \sum_{\forall b} \left[ \alpha_{1m}[b]  C_{1m}^{\mathrm{fso}}[b] +\gamma_{1m}[b]C_{1m}^{\mathrm{rf}}[b]\right]-\bar{\tau}_m\right) \nonumber\\
&+\sum_{\forall m}\lambda_{2m}\left( \dfrac{1}{B} \sum_{\forall b} \left[ \alpha_{2m}[b]  C_{2m}^{\mathrm{fso}}[b] +\gamma_{2m}[b]C_{2m}^{\mathrm{rf}}[b]\right]-\bar{\tau}_m\right),
\end{IEEEeqnarray}
}{%
\begin{IEEEeqnarray}{lll}\label{Eq:Lag}
\Scale[0.95]{
\displaystyle \mathcal{L}(\boldsymbol{\bar{\tau}},\boldsymbol{\alpha},\boldsymbol{\gamma},\boldsymbol{\bar{\lambda}})=\sum_{\forall m}\bar{\tau}_m} \\
\Scale[0.95]{ \displaystyle +\sum_{\forall m}\lambda_{1m}\bigg( \dfrac{1}{B} \sum_{\forall b} \big[ \alpha_{1m}[b]  C_{1m}^{\mathrm{fso}}[b] 
 +\gamma_{1m}[b]C_{1m}^{\mathrm{rf}}[b]\big]-\bar{\tau}_m\bigg)} \nonumber\\
\Scale[0.95]{ \displaystyle +\sum_{\forall m}\lambda_{2m}\bigg( \dfrac{1}{B} \sum_{\forall b} \big[ \alpha_{2m}[b]  C_{2m}^{\mathrm{fso}}[b] 
+\gamma_{2m}[b]C_{2m}^{\mathrm{rf}}[b]\big]-\bar{\tau}_m\bigg),} \nonumber
\end{IEEEeqnarray}
}
 where  $\boldsymbol{\bar{\lambda}}$ is a vector containing  all   Lagrange multipliers corresponding to the constraints in (\ref{Eq:Opt_DUL}), i.e., $\lambda_{lm}, \forall m,l$. The dual function and the dual problem are given by
\begin{IEEEeqnarray}{lll}
\mathcal{D}(\boldsymbol{\bar{\lambda}})=\underset{\boldsymbol{\bar{\tau}}\geq\textbf{0},\boldsymbol{\alpha}\in{\widetilde{\boldsymbol{\mathcal{A}}}},\boldsymbol{\gamma}\in{{\boldsymbol{\mathcal{G}}}}}{\mathrm{maximize}}\,\,\mathcal{L}(\boldsymbol{\bar{\tau}},\boldsymbol{\alpha},\boldsymbol{\gamma},\boldsymbol{\bar{\lambda}}) \label{Eq:Dual_F} \\
\text{and}\quad \underset{\boldsymbol{\bar{\lambda}}\geq \mathbf{0}}{\mathrm{minimize}}\,\,\mathcal{D}(\boldsymbol{\bar{\lambda}}),\label{Eq:Dual_P} 
\end{IEEEeqnarray}
respectively.
To solve (\ref{Eq:Opt_DUL}) using the dual problem in (\ref{Eq:Dual_P}), we first obtain primal variables $\boldsymbol{\bar{\tau}}$, $\boldsymbol{\alpha}$, and $\boldsymbol{\gamma}$ for a given vector of dual variables $\boldsymbol{\bar{\lambda}}$. Then, we find the dual variables  $\boldsymbol{\bar{\lambda}}$ from~(\ref{Eq:Dual_P}).

\subsection{Optimal Primal Variables}
The optimal primal variables are either  boundary points of their feasible sets  or  stationary points which can be obtained by setting  the derivatives of the Lagrangian function in (\ref{Eq:Lag}) with respect to $\boldsymbol{\bar{\tau}}$, $\boldsymbol{\alpha}$, and $\boldsymbol{\gamma}$ to zero. The derivatives of the Lagrangian function are obtained as
\begin{IEEEeqnarray}{cll}\label{Eq:Derivative}
\dfrac{\partial\mathcal{L}}{\partial\alpha_{lm}[b]}&=\dfrac{1}{B}\lambda_{lm} C_{lm}^{\mathrm{fso}}[b]  \triangleq \dfrac{1}{B} \Lambda_{lm}^{\mathrm{fso}}[b],\IEEEyesnumber\IEEEyessubnumber \\ 
\dfrac{\partial\mathcal{L}}{\partial\gamma_{lm}[b]}&=\dfrac{1}{B} \lambda_{lm} C_{lm}^{\mathrm{rf}}[b]    \triangleq \dfrac{1}{B} \Lambda_{lm}^{\mathrm{rf}}[b], \IEEEyessubnumber \\
\dfrac{\partial\mathcal{L}}{\partial\bar{\tau}_m}&=1-\lambda_{1m}-\lambda_{2m}. \IEEEyessubnumber
\end{IEEEeqnarray}
Since $\lambda_{lm} \geq 0$ holds due to dual feasibility condition \cite{Boyd}, the derivative $\dfrac{\partial\mathcal{L}}{\partial\alpha_{lm}[b]}$ in (\ref{Eq:Derivative}a) is always positive. On the other hand,  $\sum_{\forall m}\alpha_{lm}[b]=1$ has to hold for $l=1,2$. Therefore, for FSO reception, the optimal protocol selects the $\mathcal{S}-\mathcal{R}_m$ FSO link with the maximum selection metric, $\Lambda_{1m}^{\mathrm{fso}}[b]$, in each  time slot. We note that since the pdfs of the fading distributions are continuous, the probability  that two selection metrics   are equal is zero. Analogously, for  FSO transmission, the $\mathcal{R}_m-\mathcal{D}$ FSO link with the maximum $\Lambda_{2m}^{\mathrm{fso}}[b]$ will be selected. Similarly, the derivative $\dfrac{\partial\mathcal{L}}{\partial\gamma_{lm}[b]}$ in (\ref{Eq:Derivative}b) is  positive and $\sum_{\forall l} \sum_{\forall m}\gamma_{lm}[b]=1$ has to hold,  which leads to 
$\gamma_{lm}[b]=1$ for the largest $\Lambda_{lm}^{\mathrm{rf}}[b]$,  $\forall l,m$ and zero for the rest. Since $\gamma_{lm}[b]=\rho_l[b]\beta_{lm}[b]$, $\gamma_{lm}[b]=1$  leads to a unique solution for $\rho_l[b]=1$ and $\beta_{lm}[b]=1$. Moreover, since $\rho_{l}[b]=1$  holds, we obtain that $\rho_{l'}[b]=0,\,\,l'\neq l$. Therefore, the throughput does not change irrespective for which relay index $m$ $\beta_{l'm}[b]=1$ holds. Note that unique binary values are obtained for the variables of the original problem based on the optimal values of the relaxed variables. Hence, the employed relaxation also yields the optimal solution for the original problem in  (\ref{Eq:Opt_DUL}). These results are concisely stated in (\ref{Eq:OptSelec}a), (\ref{Eq:OptSelec}b), and (\ref{Eq:OptSelec}c) in Theorem~2. 

If  $\dfrac{\partial\mathcal{L}}{\partial\bar{\tau}_m} > 0$ holds, the optimal value of $\bar{\tau}_m$ is at the boundary of its feasible set, i.e., $\bar{\tau}_m \to \infty$, which cannot be the optimal solution. Similarly, if $\dfrac{\partial\mathcal{L}}{\partial\bar{\tau}_m} < 0 $ holds, the optimal value of $\bar{\tau}_m$ is at the boundary of its feasible set, i.e., $\bar{\tau}_m \to 0 $, which results in $\lambda_{1m}+\lambda_{2m} > 1 $. In addition, recall that $\lambda_{lm} \geq 0$ has to hold due to dual feasibility condition \cite{Boyd}. Therefore, either $\lambda_{1m}$ or $\lambda_{2m}$ is positive. Suppose  $\lambda_{1m}>0 \,(\lambda_{2m}>0)$ holds, then the value of  RV $\Lambda_{1m}^{\mathrm{fso}}[b] \,(\Lambda_{2m}^{\mathrm{fso}}[b])$ is greater than the value of  $\Lambda_{1m'}^{\mathrm{fso}}[b] \,(\Lambda_{2m'}^{\mathrm{fso}}[b]),\,\,\forall m'\neq m$ with a non-zero probability. Consequently, the optimal protocol will select the $\mathcal{S}-\mathcal{R}_m\, (\mathcal{R}_m-\mathcal{D})$ FSO link while the end-to-end throughput achieved by $\mathcal{R}_m$ is zero, i.e., $\bar{\tau}_m \to 0 $. This  is a contradiction. 
Therefore, the derivative $\dfrac{\partial\mathcal{L}}{\partial\bar{\tau}_m}$ in (\ref{Eq:Derivative}c) has to be zero which leads to $\lambda_{1m}+\lambda_{2m}=1$.
\subsection{Optimal Dual Variables}
 Let us first introduce a new variable  $\lambda_{m}\triangleq \lambda_{1m}=1-\lambda_{2m}$ and  vector $\boldsymbol{\lambda}$ which contains all variables $\lambda_{m},\, \forall m$. Hence, by substituting the optimal value of $\boldsymbol{\alpha}$, $\boldsymbol{\gamma}$, and $\boldsymbol{\bar{\tau}}$ into the Lagrangian function in (\ref{Eq:Lag}), the dual function in (\ref{Eq:Dual_F}) can be rewritten as 
 \iftoggle{OneColumn}{%
 \begin{IEEEeqnarray}{cll}\label{Eq:Dual_F_N}
 \mathcal{D}(\boldsymbol{\lambda})&=\sum_{\forall m}\bar{\tau}_m +\sum_{\forall m} \lambda_m\left(\bar{C}_{1m}^{\mathrm{fso}} +\bar{C}_{1m}^{\mathrm{rf}} -\bar{\tau}_m\right)+\sum_{\forall m} \left(1- \lambda_m\right) \left(\bar{C}_{2m}^{\mathrm{fso}} +\bar{C}_{2m}^{\mathrm{rf}} -\bar{\tau}_m\right)\nonumber \\
 &=\sum_{\forall m}\Big( \lambda_m\left(\bar{C}_{1m}^{\mathrm{fso}} +\bar{C}_{1m}^{\mathrm{rf}}\right) +\left(1- \lambda_m\right) \left(\bar{C}_{2m}^{\mathrm{fso}} +\bar{C}_{2m}^{\mathrm{rf}} \right)\Big),
 \end{IEEEeqnarray}
}{%
 \begin{IEEEeqnarray}{lll}\label{Eq:Dual_F_N}
 &\mathcal{D}(\boldsymbol{\lambda}) \\
 &=\sum_{\forall m}\bar{\tau}_m +\sum_{\forall m} \lambda_m\left(\bar{C}_{1m}^{\mathrm{fso}} +\bar{C}_{1m}^{\mathrm{rf}} -\bar{\tau}_m\right) \nonumber \\
 &+\sum_{\forall m} \left(1- \lambda_m\right) \left(\bar{C}_{2m}^{\mathrm{fso}} +\bar{C}_{2m}^{\mathrm{rf}} -\bar{\tau}_m\right)\nonumber \\
 &=\sum_{\forall m}\Big( \lambda_m\left(\bar{C}_{1m}^{\mathrm{fso}} +\bar{C}_{1m}^{\mathrm{rf}}\right) +\left(1- \lambda_m\right) \left(\bar{C}_{2m}^{\mathrm{fso}} +\bar{C}_{2m}^{\mathrm{rf}} \right)\Big),\nonumber
 \end{IEEEeqnarray}
}
where $\bar{C}_{lm}^{\mathrm{fso}}=\dfrac{1}{B} \sum_{\forall b}  \alpha_{lm}[b]  C_{lm}^{\mathrm{fso}}[b]$ and $\bar{C}_{lm}^{\mathrm{rf}}=\dfrac{1}{B} \sum_{\forall b}\gamma_{lm}[b]C_{lm}^{\mathrm{rf}}[b],\, l=1,2$.

The optimal value of $\boldsymbol{\lambda}$ can be obtained by solving the dual problem in (\ref{Eq:Dual_P}). In order to solve the dual problem, we use the well-known sub-gradient method~\cite{Boyd}. To minimize $\mathcal{D}(\boldsymbol{\lambda})$, the sub-gradient method updates all  component of $\boldsymbol{\lambda}$ using the following update equation in iteration $k$
\begin{IEEEeqnarray}{cll}\label{Eq:Sub_Grd}
{\lambda_m}[k+1]=\left[  {\lambda_m}[k]-\epsilon_m[k]\dfrac{\partial\mathcal{D}(\boldsymbol{\lambda})}{\partial{\lambda}_m} \right]_0^1,
 \end{IEEEeqnarray}
where $\epsilon_m[k]$ is a small step size in the $k$-th iteration. Moreover, $[\cdot]_0^1$ is used since $0\leq\lambda_m\leq 1$ has to hold. Substituting the derivative of the dual function into  (\ref{Eq:Sub_Grd}) leads to (\ref{Eq:Upd}) in Theorem~2. The results in this appendix are concisely stated in Theorem~\ref{Theo:Delay-Unlimited} which completes the proof.

\bibliographystyle{IEEEtran}
\bibliography{My_Citation_18-05-2016}

\end{document}